\documentclass[aps,pre,preprint]{revtex4-1}
\usepackage{amsthm}
\usepackage{subfigure}
\usepackage{graphicx}
\usepackage{epsfig}
\usepackage{epstopdf}
\usepackage{color}
\usepackage{caption}
\usepackage{amsmath}
\usepackage{amssymb}
\usepackage{hyperref}
\usepackage{soul}
\usepackage{wasysym}
\newtheorem{theorem}{Theorem}[section]
\theoremstyle{definition}
\newtheorem{defn}[theorem]{Definition}

\begin{document}
\title{Non-modal linear stability analysis of miscible viscous fingering in porous media}

\author{Tapan Kumar Hota, Satyajit Pramanik, Manoranjan Mishra}
\affiliation{Department of Mathematics, Indian Institute of Technology Ropar, Nangal Road, 140001 Rupnagar, India}

\begin{abstract}
The non-modal linear stability of miscible viscous fingering in a two dimensional homogeneous porous medium has been investigated. The linearized perturbed equations for Darcy's law coupled with a convection-diffusion equation is discretized using finite difference method. The resultant initial value problem is solved by fourth order Runge-Kutta method, followed by a singular value decomposition of the propagator matrix. Particular attention is given to the transient behavior rather than the long-time behavior of eigenmodes predicted by the traditional modal analysis. The transient behaviors of the response to external excitations and the response to initial conditions are studied by examining the $\epsilon-$pseudospectra structures and the largest energy growth function. With the help of non-modal stability analysis we demonstrate that at early times the displacement flow is dominated by diffusion and the perturbations decay. At later times,  when convection dominates diffusion, perturbations grow. Furthermore, we show that the dominant perturbation that experiences the maximum amplification within the linear regime lead to the transient growth. These two important features were previously unattainable in the existing linear stability methods for miscible viscous fingering. To explore the relevance of the optimal perturbation obtained from non-modal analysis, we performed direct numerical simulations using a highly accurate pseudo-spectral method. Further, a comparison of the present stability analysis with existing modal and initial value approach is also presented. It is shown that the non-modal stability results are in better agreement, than the other existing stability analyses, with those obtained from  direct numerical simulations.

\end{abstract}

\maketitle

\section{Introduction}\label{sec:introduction}

The Saffman-Taylor or viscous fingering (VF) instability, which occurs when a less viscous fluid is injected into a more viscous fluid, is a classical hydrodynamic instability phenomena \cite{Homsy1987, Saffman1958}. Such instability occurs in many physical processes such as, enhanced oil recovery\cite{Homsy1987}, spreading of surfactants coated thin liquid films \cite{Matar1999a}, liquid chromatography \cite{Mishra2008}, dispersion in aquifers \cite{Mishra2008} etc. VF  has also been used as an effective mechanism to enhance the mixing of two fluids in a micro channel \cite{Jha2011}. VF in miscible fluids were studied experimentally and theoretically by many researchers, such as Hill \cite{Hill1952}, Slobod and Thomas \cite{Slobod1963}, Perkins {\it et al.} \cite{Perkins1965}, Tan and Homsy \cite{Tan1986} are few to name. In recent times many substantial contributions \cite{Ben2002, Carvalho2014, Kim2012, Nagatsu2014,  Satyajit2013} have been made in terms of experiments, developing new mathematical tools and numerical recipes for better understanding of this phenomenon. The main mathematical challenge arises in the linear stability analysis (LSA) of miscible VF due to the unsteady base state. Thus, to analyze its LSA one needs to solve a non-autonomous system of the linearized differential operators. For the time-periodic problems one can invoke the Floquet-theory \cite{Bauer1969}. But in an arbitrary time-dependent problem, Farrell and Ioannou \cite{Farrell1996} observed that the disturbances can no longer be set proportional to $\exp(\sigma t)$, where $\sigma$ is the growth rate. As a result, the linear stability theory of the non-autonomous system prohibits to use normal mode analyses. 

The following two approaches are common in practice for analyzing the linear stability of miscible VF to find the onset of fingering: $(a)$ quasi-steady-state approximation (QSSA) and $(b)$ initial value problem (IVP). In QSSA, one reduces to an autonomous system by freezing the base state at some specific time and apply modal analysis, while in IVP approach one can find the full solution of the non-autonomous linearized problem for some representative initial conditions. Tan and Homsy \cite{Tan1986} reported that the QSSA is poorly agreeing with IVP and non-linear simulations. Although the IVP approach predicts the early time behavior better than QSSA, there are two major challenges in this method. The first one being how large the disturbances must grow before they become observable, and the other one is the choice of representative initial condition. The chosen random initial condition may not necessarily correspond to the one that gives optimum growth of the perturbation. Also, there is much disagreement on how to determine the growth of the perturbations from IVP analysis. Moreover, the random initial conditions, which are supposed to be localized within the diffusive layer, perturb the system over the entire computational domain. 

Ben {\it et al.} \cite{Ben2002} performed an LSA for VF using a spectral analysis method in a self-similar coordinate without invoking QSSA. Recently, Kim \cite{Kim2012} analyzed VF in a miscible slice and compared the predicted growth rates from QSSA with those obtained using the spectral analysis. He found that the spectral analysis predicts the system to be initially unconditionally stable and becomes unstable at later times, while in contrary, QSSA \cite{Tan1986} prediction reveals an initially unstable situation. As time progresses the predicted growth rates of the two methods get closer. A similar approach was also adopted by Pramanik and Mishra \cite{Satyajit2013} to study the effect of the Korteweg stresses on miscible VF using a self-similar QSSA (SS-QSSA). 

However, the eigenmodes obtained from both SS-QSSA and QSSA are non-orthogonal, which indicates the  possibility for transient growth of the disturbances \cite{Schmid2007}. It is a well-known fact that modal analysis does not address the phenomenon of transient growth in time \cite{Schmid2007}. In particular, it depends on the spectral properties of the underlying linear operator. For a non-normal operator (i.e., the operator that does not commute with its adjoint) an initial perturbation can grow in time by large factors before decaying, even if all the eigenmodes of the operator are damping. Earlier observations of transient growth due to the non-normality of the linearized operators were reported in the studies of instability in parallel shear flows \cite{Reddy1993, Reddy1998}, many atmospheric and laboratory flows \cite{Farrell1996, Corbett2000}. Most importantly, in all these hydrodynamic instability problems, the transient growth of infinitesimal small perturbations has been studied about a steady base state. 

In recent times, few studies \cite{Matar1999a, Shen1961, Slim2010, Doumenc2010, Rapaka2008, Rapaka2009, Bestehorn2012, Daniel2013, Tilton2013} addressed the transient growth for flows with unsteady base state. Literature pertinent {\it et al.} \cite{Rapaka2008, Rapaka2009} used non-modal analysis (NMA) to determine the optimal perturbations with maximum amplification for density driven fingering by singular value decomposition (SVD) method. Later, Doumenc {\it et al.}\cite{Doumenc2010} and Daniel {\it et al.} \cite{Daniel2013} used the `direct-adjoint looping' method \cite{Corbett2000, Schmid2007} for studying the transient growth of perturbations in Rayleigh-B\'ernard-Marogani convection and density driven fingering, respectively. Additionally, Daniel {\it et al.} \cite{Daniel2013} observed that the onset of convection in physical systems is due to suboptimal perturbations localized within the diffusive layer and hence proposed a modified optimization procedure.  

Despite the fact that transient growth analysis is known in various hydrodynamic instability problems mentioned above, to the best of authors' knowledge, the transient growth of the perturbations is yet to be investigated for miscible VF. The non-autonomous linear equations and the non-orthogonal SS-QSSA eigenmodes, which may lead to the transient growth of perturbations, motivate us to pursue the non-modal analysis in such a system. In non-modal stability theory, stability is redefined in a broader sense as the response to general input variables, including initial conditions, impulsive and continuous external excitations \cite{Schmid2007, Liu2012}. In the present paper, the transient behavior and non-normality of the linear perturbed system have been investigated in terms of responses to continuous external excitations and to initial conditions. In order to know the response to external excitations, we computed $\epsilon-$pseudospectra and the study of transient growth of the optimal perturbations is performed through the propagator matrix approach. To quantify  the maximum amplification of the perturbations we solve a matrix valued IVP, obtained from Darcy's equation coupled with a convection-diffusion equation, by fourth-order explicit Runge-Kutta method to obtain the propagator matrix. Using SVD of the propagator matrix, the singular values and the right singular vectors are obtained. The obtained singular value and right singular vector provide the optimal amplification and the optimal initial conditions, respectively. At the early times, a substantial transient growth of the perturbations is observed to exist in miscible VF, which was not shown in the existing literature \cite{Ben2002, Kim2012, Satyajit2013} of miscible VF. We also perform the direct numerical simulations (DNS) using a highly accurate Fourier pseudo-spectral method \cite{Tan1988}. The results of the amplifications obtained by non-modal theory are in good agreement with those observed from the DNS. 

The paper is organized as follows. The governing equations along with appropriate boundary and initial conditions are presented in Sec. \ref{sec:MF}. The linear perturbed equations and the numerical scheme are described in Section \ref{sec:LSA}. The rudiments of classical non-modal linear stability analysis is described in Sec. \ref{sec:TBNI}. This part also describes the non-modal approach via singular value decomposition method. Section \ref{sec:NSD} discusses our numerical findings from NMA. Section \ref{sec:DNS} contains the details about the DNS computations and non-linear growth rate. This section also discusses a comparison of previously studied linear stability methods with the results of NMA and DNS. 

\section{Mathematical Formulation}\label{sec:MF}
\begin{figure}[h!]
\centering
\includegraphics[width=3.5in, keepaspectratio=true, angle=0]
{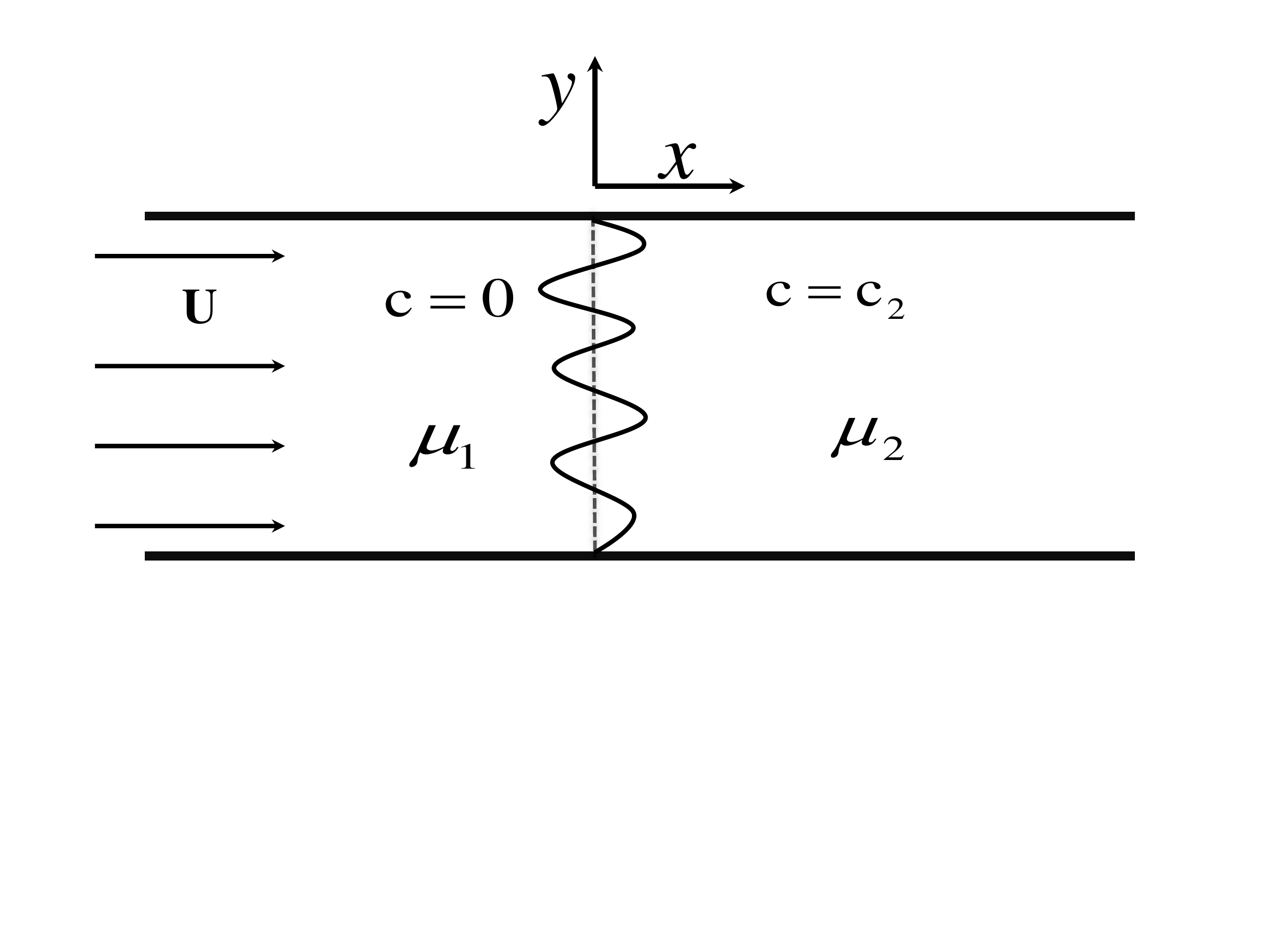} \hspace*{0cm}
\setlength{\abovecaptionskip}{-2cm}
\caption{Initially the interface is flat (dashed line) and then a wave like infinitesimal small perturbation is applied.}\label{figure_1}
\end{figure}

We consider a uniform rectilinear displacement of a fluid of viscosity $\mu_2$ by another fluid of viscosity $\mu_1$ in a homogeneous porous medium, as shown in Fig. \ref{figure_1}. The uniform displacement velocity is $Ue_x$, where $e_x$ is the unit vector along the $x-$direction. The viscosity of the two fluids is determined in terms of a solute concentration, $c$, which is assumed to be $c = 0$ and $c = c_2$ in the displacing and displaced fluids, respectively. Fluids are assumed to be non-reactive, neutrally buoyant, and incompressible. The porous media have constant permeability, $\kappa$, and an isotropic dispersion $D$.

\subsection{Governing Equations}\label{subsec:GE}
The above-mentioned displacement flow in porous media can be described by the Dacry's law for the fluid velocity coupled with a convection-diffusion equation that determines the evolution of the solute concentration. For the non-dimensional formulation of the problem we use $U, ~ D/U, ~ D/U^2, ~ \mu_1, ~ \mu_1D/\kappa$ as the characteristic velocity, length, time, viscosity, and pressure, respectively. The related non-dimensional equations in a reference frame moving with velocity $Ue_x$ are \cite{Tan1986} 
\begin{eqnarray} 
\label{cont_eqn}
& & \nabla \cdot \underline{u} = 0,\\ 
\label{Darcy_1}
& & \nabla p = -\mu(c) (\underline{u} + e_x), \\ 
\label{convec_diffuse}
& & \frac{\partial c} {\partial t} + \underline{u} \cdot \nabla c = \nabla^2 c, 
\end{eqnarray}
where $\underline{u} = (u, v)$ is the two-dimensional Darcy velocity. The only parameter that enters the dimensionless viscosity-concentration relation is the log-mobility ratio $R$, defined as, $R= \ln (\mu_2 / \mu_1)$, and the viscosity is related to the concentration by an Arrhenius type relationship \cite{Tan1986}, i.e., $\mu(c) = \exp(Rc)$. The coupled Eqs. \eqref{cont_eqn} -- \eqref{convec_diffuse} are associated with following initial and boundary conditions \cite{Nield1992}. 

Initial conditions: 
\begin{eqnarray}\label{IC_1}
\underline{u}=(0,0), ~~~
c(x,y,t=0)=\begin{cases}
0, & x \leq 0\\
1, & x > 0,
\end{cases} ~~~~\forall y. 
\end{eqnarray}  

Boundary conditions:
\begin{eqnarray}
\label{bc_1}
& & \underline{u} = (0, 0), ~~~ \frac{\partial c}{\partial x} =0, ~~~ |x| \rightarrow \infty, ~~~ \text{streamwise direction}, \\
\label{bc_2}
& & \frac{\partial c}{\partial y} = 0, ~~~ \frac{\partial v}{\partial y} = 0, ~~~ \forall x, ~~~ \text{spanwise direction}.
\end{eqnarray} 
 
\subsection{Base State}\label{subsec:BS}
The base state of the flow is assumed to be pure diffusion of
the concentration along the axial direction (i.e., no advection $ \underline{u} = (0,0)$). Hence, the time-dependent base-state solution of Eqs. \eqref{cont_eqn}-\eqref{convec_diffuse} on infinite domain with the no-flux boundary conditions for the concentration is \cite{Tan1986} 
\begin{eqnarray}
\label{base_state}
\underline{u}_0=(u_0, v_0)&=&\underline{0},\\
\mu_0=\mu_0(c_0)&=&\mu_0(x,t),\\
p_0(x,t)&=&-\int_{-\infty}^{x} \mu_0(s,t)ds,  \\
c_0(x,t)&=&\frac{1}{2}[1+\text{erf} (x/2\sqrt{t})],
\label{concen_base_state}
\end{eqnarray}
where $\text{erf}(z)= \frac{2}{\sqrt{\pi}} \int_{0}^{z} e^{-t^2} \mbox{d}t$ is the error function. On transforming the current coordinates $(x,y,t)$ to a new coordinate system $(\xi,y,t)$, the base state concentration (Eq. \eqref{concen_base_state}) becomes 
\begin{equation}\label{concen_xi_t}
\displaystyle c_0(\xi) = \frac{1}{2}\left[1+\text{erf} (\xi/2)\right],
\end{equation}
where  $\xi:= x/ \sqrt{t}$ is the similarity variable transformation. Advantage of working with $(\xi,y, t)$ coordinate systems is that, one can pretend that as if the base state is time independent.

\section{Linear Perturbation Equations}\label{sec:LSA}
In this section, we discuss the method of numerical solutions of the linearized equations. The linear perturbation equations are solved as an IVP in both $(x,t)$ and $(\xi,t)$ coordinate systems, and the obtained numerical results are compared. The response of the system to small disturbances is obtained by linearizing Eqs. \eqref{cont_eqn}-\eqref{convec_diffuse} according to  
\begin{equation}
u(\xi, y, t) = u'(\xi,y,t),\;c(\xi,y,t) = c_0(\xi) + c'(\xi,y,t), 
\end{equation}
where $u'(\xi,y,t), ~ c'(\xi,y,t)$ represent infinitesimal disturbances to the velocity and concentration, respectively. We seek disturbances of the form $(u',c')(\xi,y,t) = (u', c')(\xi,t) e^{iky}$, describing a function which evolves in time and in the streamwise direction while exhibiting sinusoidal character in the transverse direction. Using viscosity-concentration relationship $\mu(c)= e^{Rc}$, the disturbances $u'$ and $c'$ are determined from the following set of equations representing a disturbance of transverse wave number $k$ \cite{Satyajit2013}, 
\begin{eqnarray}  
\label{Linearized_31}
& & \left[ \mathcal{D} + \frac{R}{2 \sqrt{\pi}}\exp\left(\frac{-\xi^2}{4}\right) \frac{\partial}{\partial \xi} \right] u'(\xi,t) = k^2 R t c'(\xi, t),\\
\label{Linearized_32} 
& & \left[ t\frac{\partial}{\partial t}  - \frac{\xi}{2} \frac{\partial}{\partial \xi}- \mathcal{D}\right] c'(\xi,t)= -\frac{\sqrt{t}}{2 \sqrt{\pi}} \exp\left(\frac{-\xi^2}{4}\right) u'(\xi,t), 
\end{eqnarray} 
where $\mathcal{D}:= \partial^2/\partial \xi^2 - tk^2$. The associated boundary conditions are  $(c', u') \rightarrow (0, 0),$ as $ |\xi| \rightarrow \infty$. Note that $u'$ and $c'$ are not periodic in the `$\xi$' variable due to the far field conditions i.e., $(c', u') \rightarrow (0, 0), ~ |\xi| \rightarrow \infty$. So the functions $u'$ and $c'$ cannot be represented as a superposition of sinusoidal perturbations in the `$\xi$' variable. The analytic solution for the coupled system of partial differential Eqs. \eqref{Linearized_31} and \eqref{Linearized_32}, for a diffusive base state is unattainable. Hence, a numerical solution method needs to be used for solving these coupled equations. In this paper we use finite difference method and solved the linear stability problem in a finite computational domain.

\begin{figure}
\centering
\includegraphics[width=3.2in, keepaspectratio=true, angle=0]{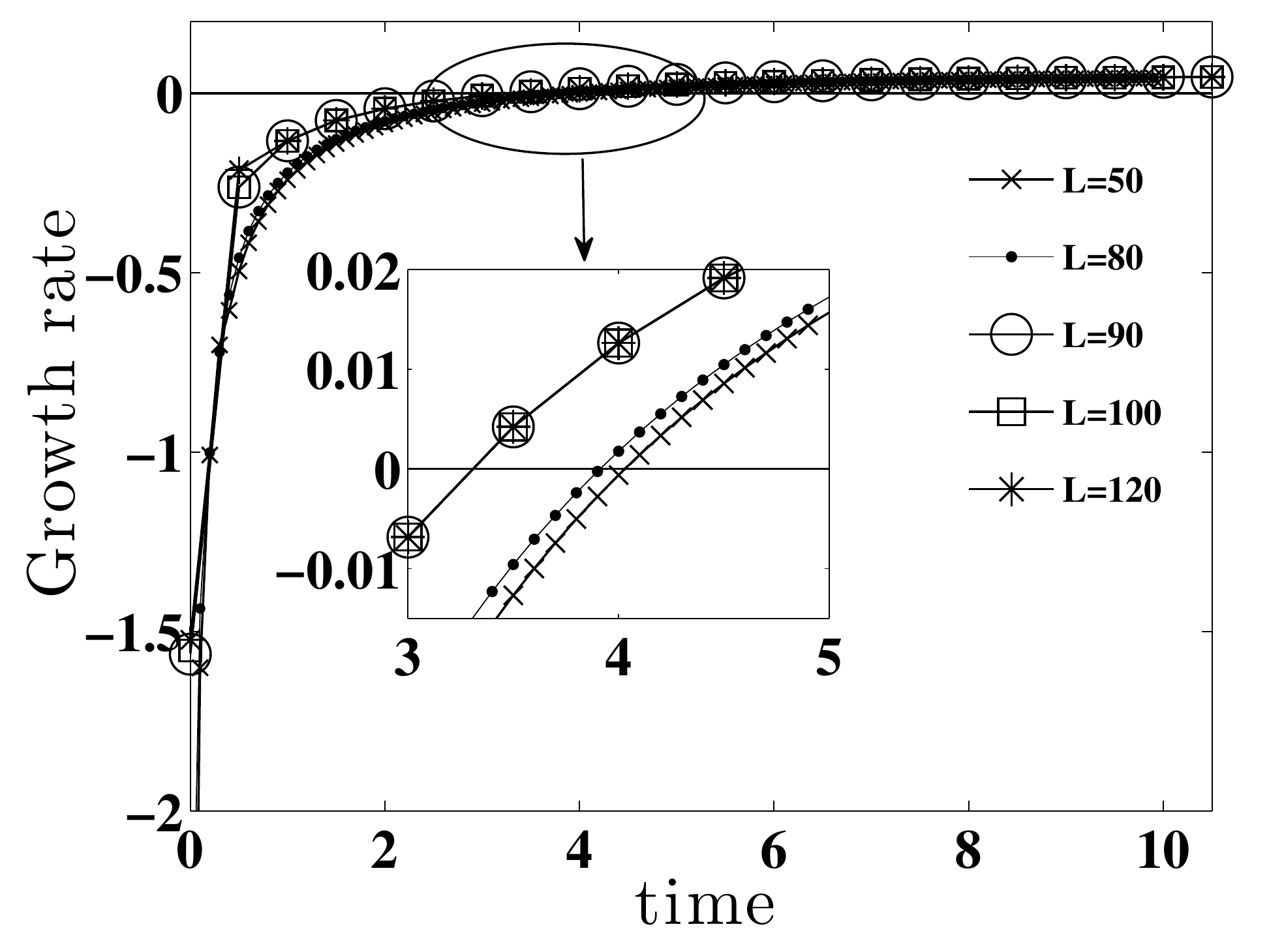}
\setlength{\abovecaptionskip}{-.2cm}
\caption{Dependence on the size of the computational domain on growth rate obtained from NMA, for $R=3, \; k=0.2$. Beyond the domain length $L=90$, there are  almost no change.}
\label{domain_length}
\end{figure}

\subsection{Numerical Method}\label{subsec:NS}
For a given $k$ and $R$, we discretize Eqs. \eqref{Linearized_31} and \eqref{Linearized_32} over a finite computational domain $[-L, L]$, $L \in \mathbb{R}^+$, with a uniform mesh size $h$ and $n + 2$ number of grid points. All the derivatives are approximated by central difference formulae. The discretized version of  Eqs. \eqref{Linearized_31} and \eqref{Linearized_32} with the boundary conditions $u'(t)\Bigm|_{i=0} = 0 = c'(t)\Bigm|_{i=0}$ and  $u'(t)\Bigm|_{i=n+1}=0,c'(t)\Bigm|_{i=n+1}=1$ can be cast into a non-autonomous system
\begin{equation}\label{IVP_1}
\frac{\text{d} c'(t)}{\text{d} t}= \underline{A}(t) c',
\end{equation}
where $c'_{i}(t) = c'(\xi_i,t)$ and $\underline{A}(t)= M_3+M_4M_1^{-1}M_2$ is the stability matrix of order $n$. Here each matrix $M_j, ~ j=1, 2, 3, 4$ is  either a diagonal matrix or a tridiagonal matrix and are given by \cite{Satyajit2013},
\[M_1(i,j)=\begin{cases}
\frac{1}{h^2} \pm \frac{R}{4 h^2} \left(  c_0(j+1)-c_0(j-1)\right), & i = j \mp 1 \\
-\frac{2}{h^2}-k^2 t, & i = j \\
0, & \text{otherwise,}
\end{cases}\]

\[M_2(i,j) = \begin{cases}k^2 R t, & i=j\\
0, & \text{otherwise,} \end{cases}\]

\[M_3(i,j)=\begin{cases}
\frac{1}{th^2} \pm \frac{\xi_j}{4ht}, & i = j \mp 1 \\
-\frac{2}{t h^2}-k^2, & i = j \\
0, & \text{otherwise,}
\end{cases}\]
and 
\[M_4(i,j)= \begin{cases}- \left(\frac{c_0(j+1)- c_0(j-1)}{2 h\sqrt{t}}\right), & i=j\\
0, & \text{otherwise.} \end{cases}\]

For the computational purpose the unstable interface is set at $\xi=0$. The initial value problem, Eq. \eqref{IVP_1}, is solved using Runge-Kutta fourth-order explicit method for time integration with a relative error of order $\mathcal{O}(10^{-10})$. The matlab built-in functions \textsf{ode45} and \textsf{eigs} are used for solving the IVP and finding the eigenvalues, respectively.

The computational domain is taken sufficiently large so that the numerical results remain unaffected by the truncation of infinite physical domain into finite domain in numerical algorithms. To ensure that our numerical quantities, \textit{viz.} singular values, eigenvalues, etc., that characterize the instability, are independent of spatial domain a series of numerical calculations are performed for different $L$ ranging from $50$ to $120$. The parameter values are chosen to be $R = 3, k = 0.2$ \cite{Tan1986}. The obtained numerical results are compared in Fig.\ref{domain_length}, and it is observed that for $L$ $\geq$ 90 there is no change in the growth rates calculated from NMA which is explained in Sec. \ref{sec:NMA}. Thus for optimal result, the computational domain is chosen $[-100, 100]$ in all our calculations.

\begin{figure*}[!ht]
\centering
(a) \hspace{8cm} (b)\\
\includegraphics[width=3.2in, keepaspectratio=true, angle=0]{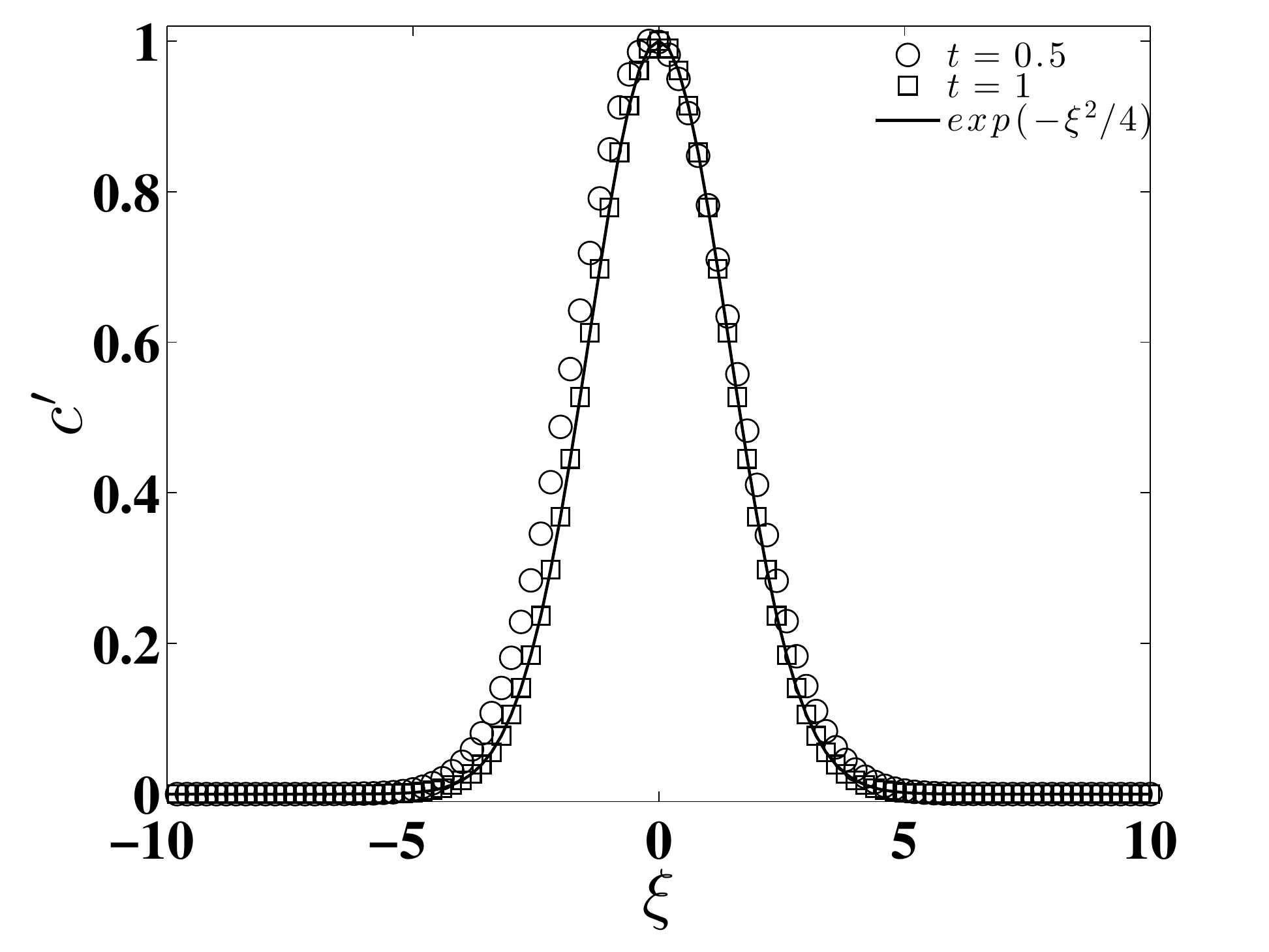}
\includegraphics[width=3.2in, keepaspectratio=true, angle=0]{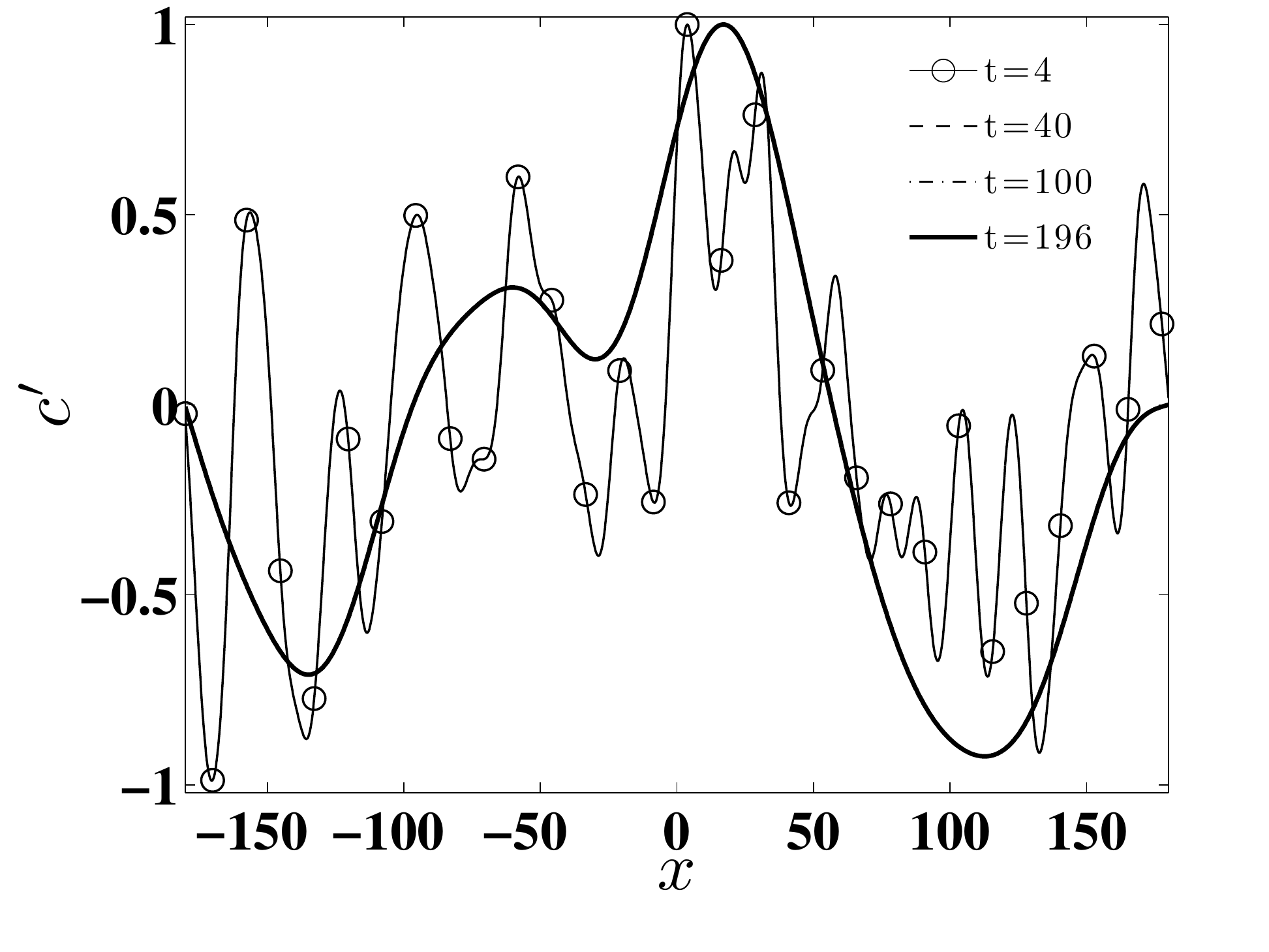}
\caption{Disturbance concentration profile for $R=3, k=0$ at different times, obtained from the IVP in $(a)$ $(\xi, t)$ coordinate system, $(b)$ $(x, t)$ coordinate system.}
\label{xi_t_IVP_1}
\end{figure*}

Numerical convergence of finite difference method has been tested with different number of grid points, and the optimal computational time and accuracy were obtained with the spatial step size $0.2$. As a verification of our numerical code we reproduce the results of Rapaka  \textit{et al.} \cite{Rapaka2008}. Due to the possible singularity at $t=0$, all the numerical integrations are performed starting from the initial time $ t = 10^{-3}$ to various final time. Although we restrict our study of the optimal perturbations and maximum amplification for $R = 3$, the analogous transient growth dynamics of perturbations can be observed for other positive values of $R$. The next section (Sec. \ref{subsec:IVP}) describes the importance of NMA for the present problem by using the non-autonomous system (Eq. \eqref{IVP_1}) and explains the result of IVP calculations in $(x,t)$ and $(\xi,t)$ co-ordinate systems. 

\subsection{Initial Value Calculation}\label{subsec:IVP}
In this section, we illustrate the advantages of studying stability analysis in $(\xi,t)$ domain. For this, we solve the non-autonomous system  Eq. \eqref{IVP_1} with respect to the following random initial condition 
\begin{eqnarray}\label{random_IC}
c'(\omega,t_0)= \delta * \text{rand}(\omega),~ \text{where}~ \omega= \xi ~\text{or}~ x, 
\end{eqnarray}
where $\text{rand}(\omega)$ represents a random number generator between $-1$ and $1$, and an infinitesimal parameter $\delta$ determines the amplitude of the perturbation, which is chosen to be $\mathcal{O}(10^{-3})$. The objective of performing IVP in both the coordinate systems is to describe the structure of perturbed concentration in each coordinate system. The initial time $t_0=10^{-3}$ has been chosen for comparing the structure of perturbed concentration obtained from $(x,t)$ and $(\xi, t)$  coordinates. Fig. \ref{xi_t_IVP_1}(a) depicts that the concentration eigen-function determined by solving the IVP in $(\xi,t)$ domain converges rapidly to the dominant eigenmode $\exp(-\xi^2/4)$ \cite{Ben2002}. On the other hand, Fig. \ref{xi_t_IVP_1}(b) illustrates that the concentration eigen-function obtained from the IVP calculation in $(x,t)$ coordinate system takes longer time to converge to the dominant mode, in comparison to calculations in $(\xi,t)$ coordinate system. Both the results shown in Fig. \ref{xi_t_IVP_1} are obtained by averaging over $10$ random initial conditions. The localized eigen-functions in $(\xi,t)$ coordinates converge rapidly to the exact solution, thereby giving an accurate disturbance growth rate at small times. With this observation, rest of the numerical calculations are performed in the self-similar $(\xi,t)$ domain. We perform IVP calculations in $(\xi,t)$ domain with dominant eigenmode $\exp(-\xi^2/4)$ as the initial condition and the obtained results are discussed in section \ref{Dispersion_Growrthrate}.

\section{Transient Behaviors and Non-modal Instabilities}\label{sec:TBNI}
In this section, the stability analysis is carried in a broad sense as the responses to external excitations and to the initial conditions. We describe both of these through rigorous mathematical analysis. Suppose  the system is subjected to a  perturbation of the form
\begin{equation}\label{exponent_perturb}
u'(\xi, t)= u'(\xi)\exp(\sigma(t_0) t),\; c'(\xi, t)= c'(\xi)\exp(\sigma(t_0) t), ,
\end{equation}
where $\sigma$ is the complex frequency, and $t_0$ is the diffusive time at which the base state is frozen. Thus Eq. \eqref{IVP_1} is converted into an eigenvalue problem $\underline{A}(t_0)c'= \sigma(t_0) c'$. To analyse the system behavior to the external excitation the study of $\epsilon-$pseudospectra\cite{Schmid2007, Liu2012, Trefethen2005} is a very helpful tool. For given $\epsilon >0$, the $\epsilon-$pseudospectra is defined as
\begin{equation}\label{epsilon-pseudo}
\Lambda_\epsilon(\underline{A}) = \{z \in \mathbb{C} : \parallel R_z(\underline{A}) \parallel \geq \epsilon^{-1}\},
\end{equation}
where $R_z(\underline{A})= \left( z \mathcal{I} - \underline{A}\right)^{-1}$ is known as the resolvent, $\mathcal{I}$ is the identity operator, and $\parallel \cdot \parallel$ is the Euclidean $2-$norm in $\mathbb{C}^n$.

In pursuit of optimal amplification of initial conditions, we follow the classical propagator matrix approach. The maximal amplification is obtained as the singular value decomposition of this propagator matrix and the detail analysis is given in Sec. \ref{sec:NMA}. This section also describes the choice of measure in terms of Euclidean $2-$norm. Often for IVP in  hydrodynamics, one may be interested in the dynamics of the energy growth rate, for  $t \ll 1$. In this context, the concept of numerical range can be used to link the stability matrix to the early time energy growth \cite{Schmid2007, Trefethen2005}. The numerical range is defined as
\begin{eqnarray}\label{numerical_range}
W(\underline{A})= \{\textbf{x}^*\underline{A} \textbf{x}: \textbf{x} \in \mathbb{C}^n, \parallel \textbf{x} \parallel = 1\},
\end{eqnarray}
where the superscript $*$ denotes the transpose. The growth or decay of the initial energy can be studied from the numerical abscissa
\begin{equation}\label{numerical_abscissa}
\eta(\underline{A}) = \displaystyle\ \sup_{z \in W(\underline{A})} \Re (z) = \sup \{\lambda(\underline{A} +\underline{A}^*)/2\},
\end{equation}
where  $\lambda(\cdot)$ represents the spectrum of the matrix and $\Re(\cdot)$ denotes the real part. The numerical abscissa $\eta(\mathcal{A})$ measures the maximum possible instantaneous growth rate corresponding to any initial condition as $t \to 0$ \cite{Trefethen2005}.

\begin{figure*}
\centering
(a) \hspace{5cm} (b)\\
\includegraphics[width=2in, keepaspectratio=true, angle=0]{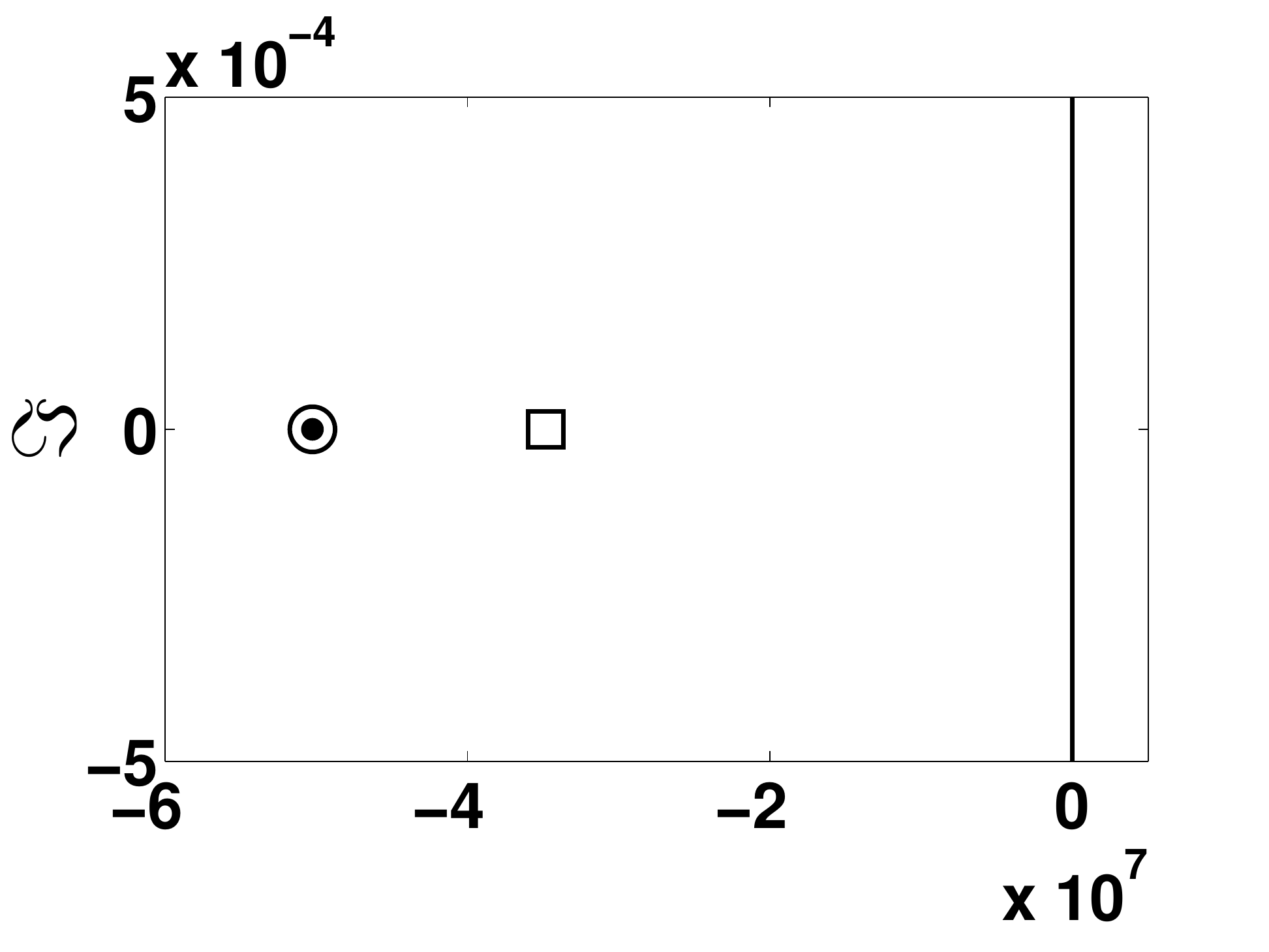}
\includegraphics[width=2in, keepaspectratio=true, angle=0]{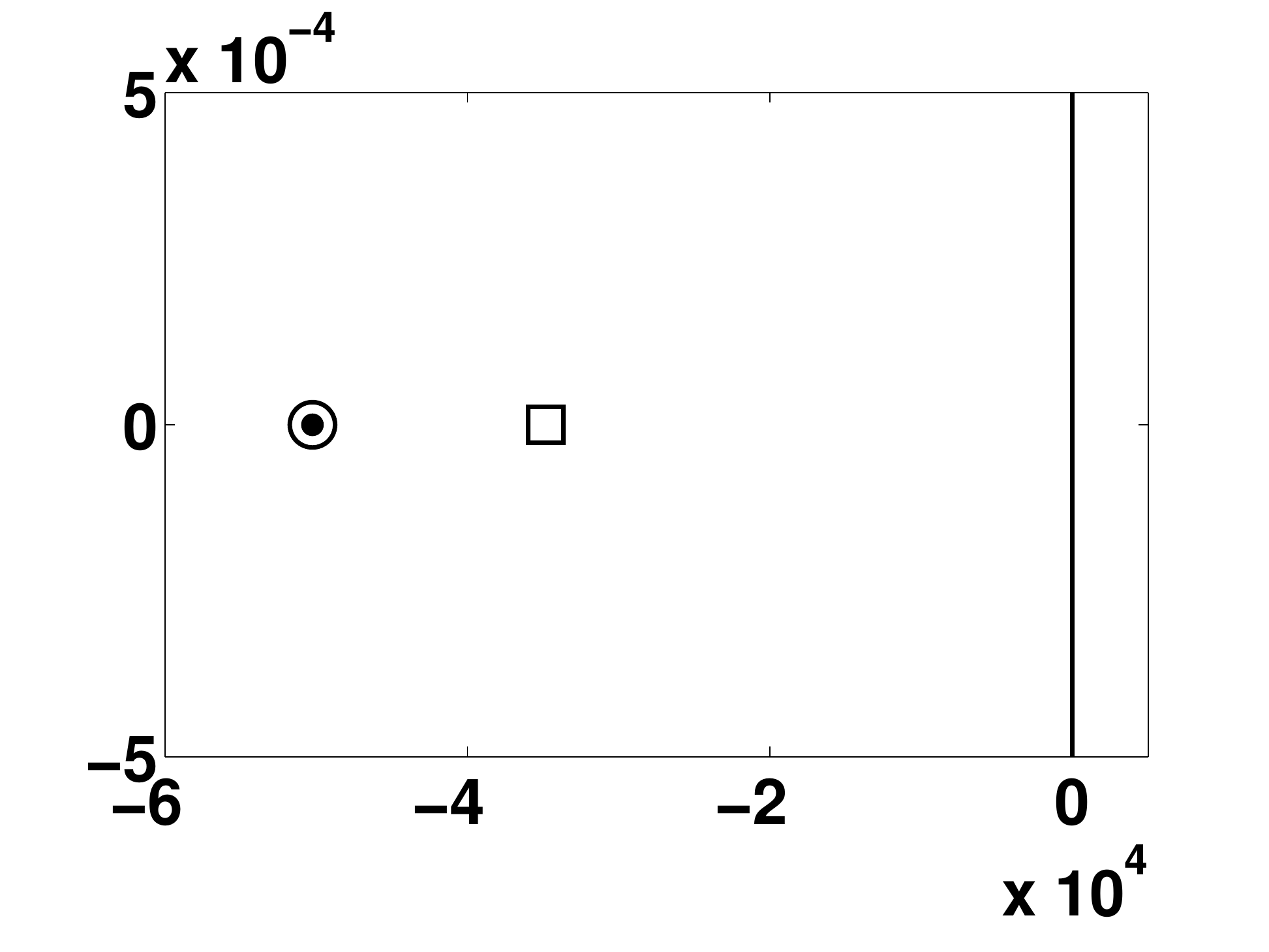}\\
(c) \hspace{5cm} (d)\\
\includegraphics[width=2in, keepaspectratio=true, angle=0]{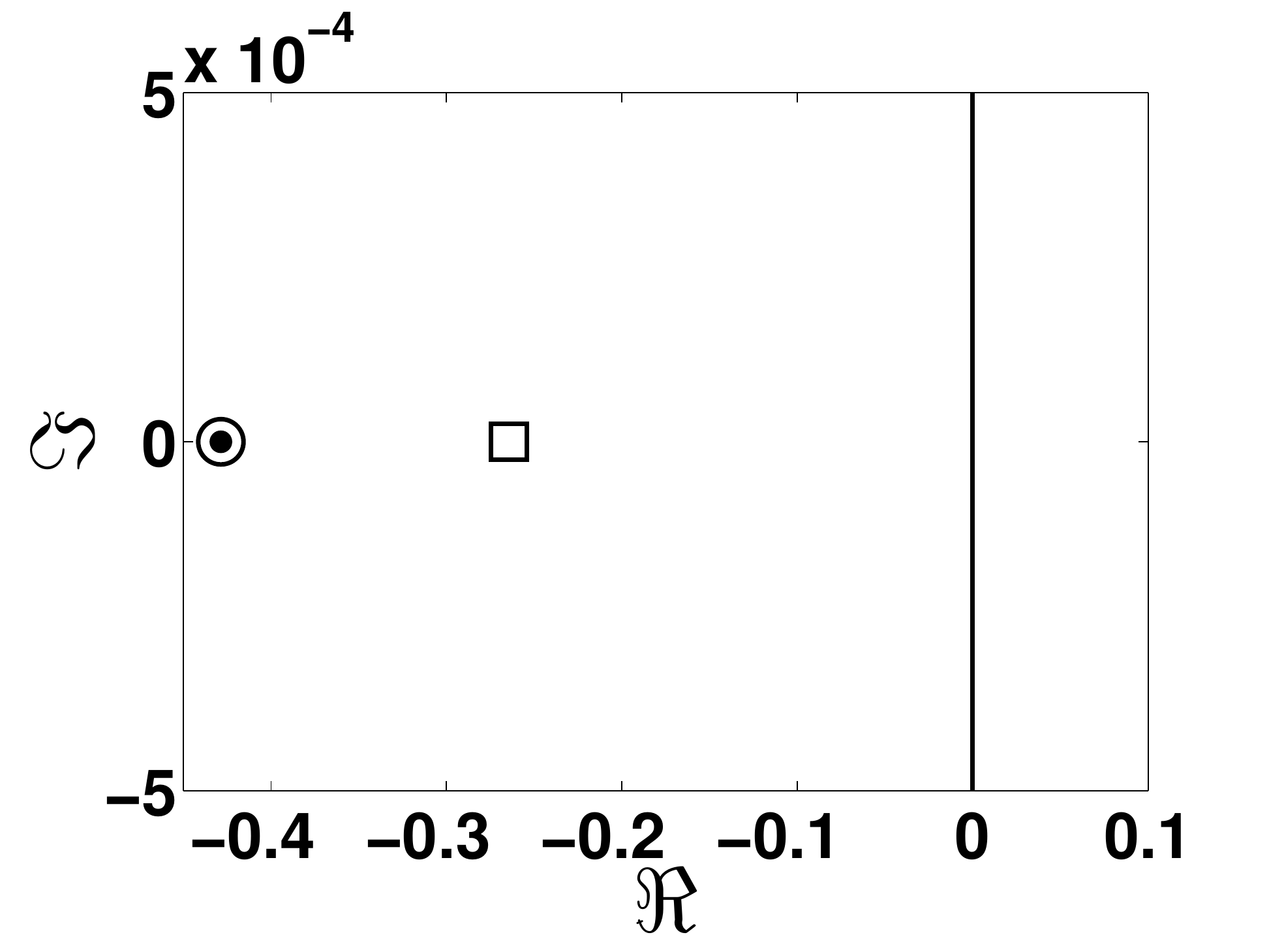}
\includegraphics[width=2in, keepaspectratio=true, angle=0]{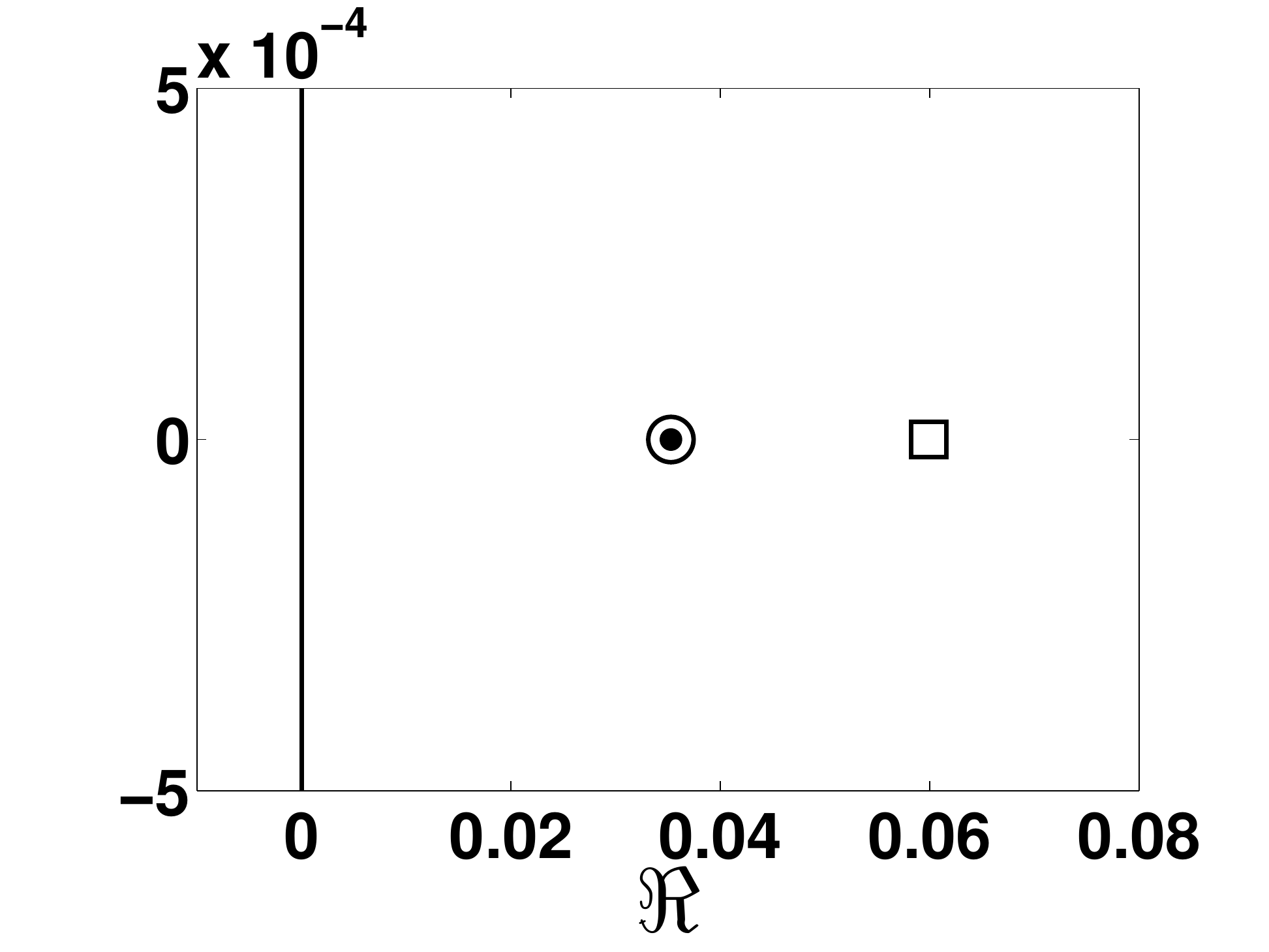}\\
\setlength{\abovecaptionskip}{0cm}
 \caption{Spectral abscissa (circles) and numerical abscissa
 (squares) for $k=0.2, \:R=3$, at different time $(a)~ t = 10^{-8}, (b) ~t = 10^{-5}, (c)~ t = 1, (d) ~t = 10$ . Also shown the eigenvalue with largest real part(black dot) and imaginary axis as bold continuous vertical line. $\Re$ and  $\Im$ denote the real and imaginary axis in the complex plane.} 
\label{abscissa}
\end{figure*}

\subsection{Pseudospectra}\label{subsec:Pseudo}
The behavior of the eigenvalues of the linearized stability matrix $\underline{A}(t)$ determine the instability of the system. But, for the highly non-normal matrices, e.g., in bounded shear flows \cite{Schmid2007, Reddy1993}, the classical normal mode analysis, which assumes an exponential time dependence of $\underline{A}(t)$, fails to predict the instability appropriately. The temporal eigenmodes of the time dependent matrix $\underline{A}(t)$ are not defined \cite{Farrell1996} and to determine the asymptotic behavior (by the Lyapunov exponents) of Eq. \eqref{IVP_1} is beyond the scope of this paper. To quantify the non-normality of $\underline{A}(t)$, we freeze $\underline{A}(t)$ at different times and consider two quantities namely the spectral abscissa and the numerical abscissa.
 The spectral abscissa of the stability matrix $\underline{A}$ is defined as
\begin{eqnarray}\label{spectral_abscissa}
\alpha(\underline{A}) &:=& \displaystyle{\max \{\Re(\lambda(\underline{A}))\}}
\end{eqnarray}
It can be verified that for a normal matrix $\alpha(\underline{A}) =\eta(\underline{A})$. The main objective of modal analysis is to study the spectral abscissa and the corresponding eigenmodes. In Fig. \ref{abscissa}, the spectral and numerical abscissa for the stability matrix $\underline{A}$ are plotted at different time \st{with} for given wave number $k=0.2$ and $R=3$. It is seen from Figs. \ref{abscissa}(a) and \ref{abscissa}(b) that at early times, the spectral abscissa and the numerical abscissa differ from each other in the order of $\mathcal{O}(10^{7})$ and  $\mathcal{O}(10^{4})$. This reveals that at early time, the stability matrix $\underline{A}$ is highly non-normal, but at later times, $\underline{A}(t)$ is close to a normal matrix (see Figs. \ref{abscissa}(c) and \ref{abscissa}(d)).  

\begin{figure*}
\centering
(a) \hspace{8cm} (b)\\
\includegraphics[width=3.2in, keepaspectratio=true, angle=0]{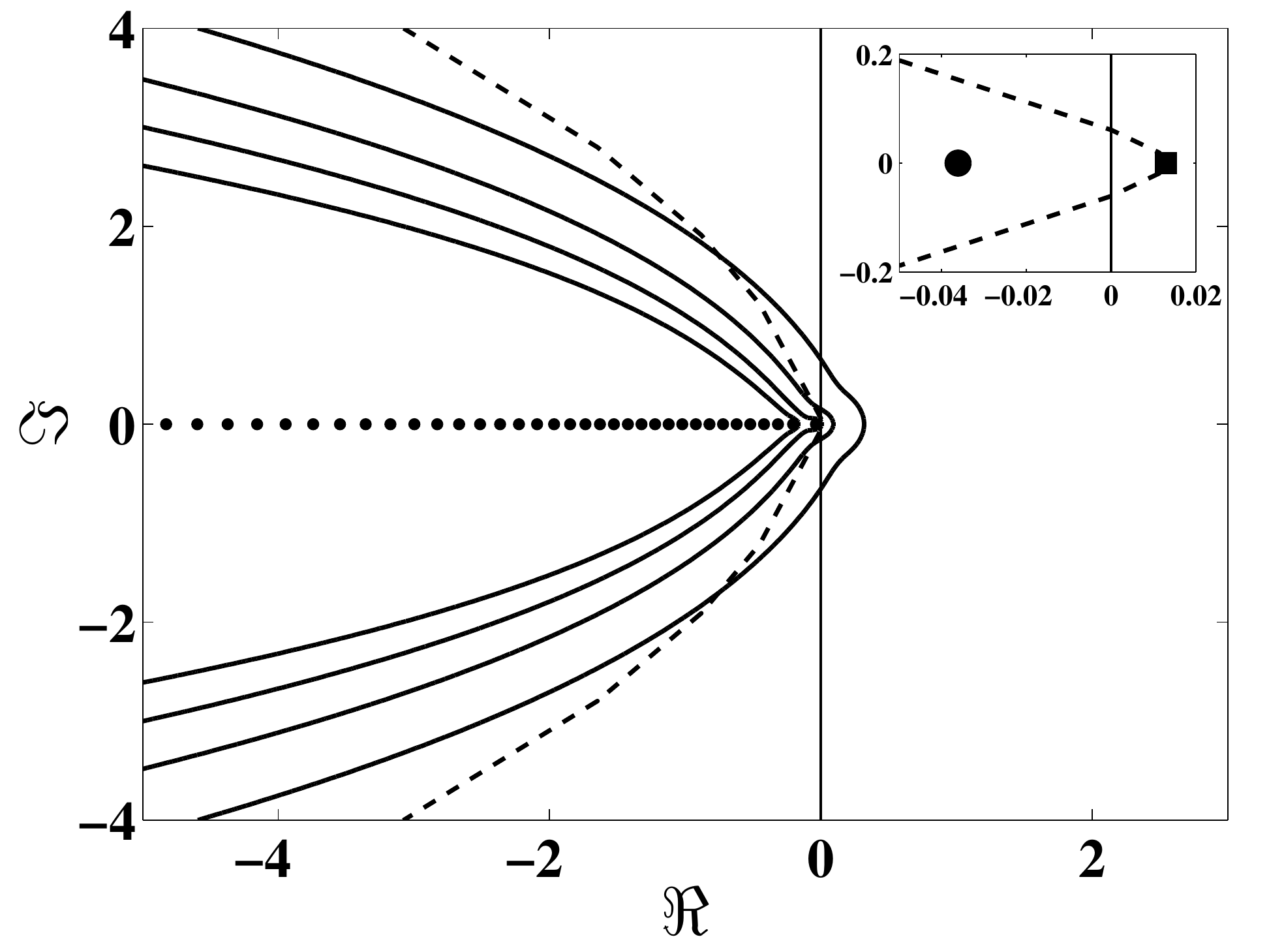}
\includegraphics[width=3.2in, keepaspectratio=true, angle=0]{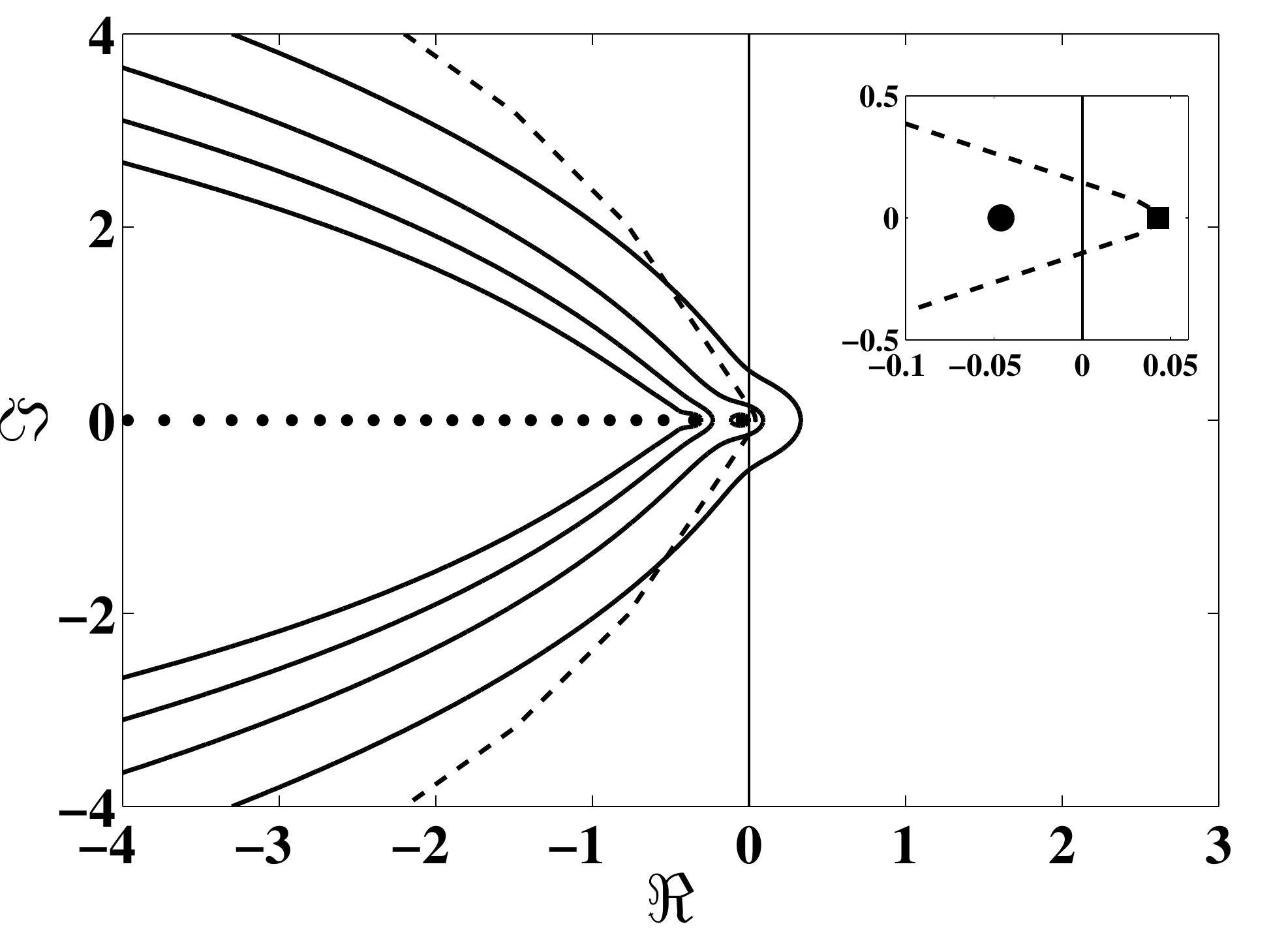}
\caption{Pseudospectra of $\underline{A}(t)$ for (a) $R = 2, k = 0.15, t = 5$, (b) $R = 3, k = 0.25, t = 3$. Black square: numerical abscissa; black dots: eigenvalues; dashed line: boundary of the numerical range; solid lines: contours from innermost to outermost representing levels from $\epsilon = 10^{-2}$ to $10^{-0.5}$ with increment $10^{0.5}$. Here $\Re$ and $\Im$ represent the real and imaginary axis in the complex plane.}
\label{fig:epsilon_pseudo}
\end{figure*}

For the unsteady base state, Pramanik \textit{et.al} studied the response to external excitations by analysing the pseudospectrum \cite{Satyajit2015}. Following these authors, we plot $\epsilon-$pseudospectra using Eigtool \cite{Thomas2002}.  Fig. \ref{fig:epsilon_pseudo} illustrates the pseudospectra for $R = 2, k = 0.15, t=5$ and $R=3, k=0.25, t=3$. The contour lines indicate isolines of constant resolvent norm. We observe that the numerical range partially reaches into the unstable half-plane and it strictly contains the spectrum. This means that there exists positive energy growth rates, despite the fact that the spectrum is contained inside the stable half-plane. This also shows that the eigenvalues are inherently time-asymptotic tools when dealing with stability analysis of non-normal matrices. We thus conclude that initial energy growth is possible up to a growth rate given by the maximum protrusion of the numerical range into the unstable half-plane and the least stable eigenvalue governs the long time behavior. This early time structure of non-normality of $\underline{A}(t)$ inspires us to investigate the transient growth of perturbations and the onset of instability, which is presented in Sec. \ref{sec:NMA}.


\subsection{Transient energy growth}\label{sec:NMA}
In this section, we present the mathematical analysis for calculating the transient energy growth and the onset of instability. Let us assume a solution of Eq. \eqref{IVP_1} as 
\begin{equation}\label{Propagator}
c'(t):= \Phi(t_0; t) c'_0,
\end{equation}
where $c'(t_0)=c'_0$ is an arbitrary initial condition. Here $\Phi(t_0;t)$ is called the propagator matrix, because it propagates the information forward from the initial time $t_0$ to time $t$. On substituting Eq. \eqref{Propagator} into Eq. \eqref{IVP_1} we obtain a matrix differential equation,
\begin{equation}\label{IVP_Matrix}
\frac{\mbox{d}}{\mbox{d}t} \Phi(t_0; t)= \underline{A}(t) \Phi(t_0;t),
\end{equation}
with the initial condition $\Phi(t_0;t_0)=I$, where $I$ is the identity matrix of the same order of the matrix $\underline{A}(t)$. Thus, instead of solving a vector differential equation with random initial condition, now it needs to solve a matrix differential equation, Eq. \eqref{IVP_Matrix}, with deterministic initial condition, i.e.,
$\Phi(t_0;t_0)=I$. 

The stability analysis is performed based on the amplification magnitude of the perturbations over a prescribed finite time interval \cite{Schmid2007}. In the stability analysis of a physical system with time dependent base state, the growth or decay of disturbances is only meaningful in reference to the evolved base state. In $1961$, Shen \cite{Shen1961} observed in his study of unsteady parallel shear flow that if both the disturbance and the base state are decaying, but the latter one with faster rate, the disturbance, relative to the basic state, would appear to be amplified at a later instant. Conversely, if a disturbance grows in time, but the base state grows faster than that, then the disturbance will appear to decay in time. So it is necessary that the growth or decay of a disturbance should be measured only by a comparison with the basic flow. With these observations Shen \cite{Shen1961} introduced ``momentary stability'' and defined an appropriate measure for stability analysis. Similar measures have also been adopted by Matar and Troian \cite{Matar1999a} in their studies of  spreading  of surfactants on a thin film, and Bestehorn and Firoozabadi \cite{Bestehorn2012} in their study of the dissolution of $\text{CO}_2$ in saline aquifers. 

Following the work of Shen \cite{Shen1961}, we would like to introduce the quantities such as amplification magnitude and transient growth rate of the disturbances that will be used to measure the time dependent growth rate of the perturbations. The perturbation magnitude  for the flow variable $f$ can be defined as 
\begin{equation}\label{enegry}
\displaystyle M_f(t):= \frac{E_{f}(t)}{E_{f_0}(t)}, 
\end{equation}
\noindent where $f_0 = c_0, \text{or}\;u_0$, and $f = c', \text{or}\; u'$ and $E_{f}(t) := \|f(t)\|_2^2= \int_{-\infty}^{\infty} f^2(x,t) \text{d} x,\; E_{f_0}(t) := \|f_0(t)\|_2^2= \int_{-\infty}^{\infty} f_0^2(x,t) \text{d} x$. The sensitivity of the infinitesimal disturbances introduced at time $t_0$ is measured from the ratio of the perturbation magnitude $M_f(t)$ at time $t$ to its initial value $M_f(t_0)$. Consider the normalized amplification $\Psi_f (t)$ defined by

\begin{equation}\label{Amplification}
\Psi_{f}(t) := \frac{M_f(t)}{M_f(t_0)} = \left[\frac{E_{f}(t)}{E_{f}(t_0)}\right] \big/ \left[ \frac{E_{f_0}(t)}{E_{f_0}(t_0)}\right] = \frac{G_{f}(t)}{G_{f_0}(t)}.
\end{equation}
The time-dependent growth rate is defined as \cite{Matar1999a, Bestehorn2012}

\begin{equation}\label{Growthrate}
\sigma(t) = \frac{1}{\Psi_{f}(t)} \frac{\text{d} \Psi_{f}(t)}{\text{d} t} = \frac{1}{G_{f}} \frac{\text{d} G_{f}}{\text{d} t} 
- \frac{1}{G_{f_0}} \frac{\text{d} G_{f_0}}{\text{d} t} = \sigma_f(t) - \sigma_{f_0}(t).
\end{equation}
As $c_0$ in $(\xi,t)$ coordinate is independent of time and $u_0 = 0$, this implies  $\sigma_{f_0}(t)=0$, and Eq. \eqref{Growthrate} reduces to 
\begin{equation}\label{Gr_001}
\sigma(t) = \frac{1}{G_f}\frac{\text{d}G_f}{\text{d}t}.
\end{equation}

To measure the transient growth, we consider the concentration perturbation magnitude, i.e. $f = c'$ at time $t$  which is normalized to the concentration perturbation magnitude at $t_0$. As the perturbation equations are linear, without loss of generality it can be assumed that $\| c'(t_0) \| = 1$ and using Eq. \eqref{Propagator},  Eq. \eqref{Amplification} reduces to
\begin{eqnarray}
\Psi_{c'}(t) & = & E_{c'}(t)/ E_{c'}(t_0) = \| c'(x,t) \|^2 \nonumber \\
& = & \langle \Phi(t_0; t)^* \Phi(t_0; t) c', c' \rangle,
\end{eqnarray}
where $\langle \cdot, \cdot \rangle$ is the standard $L^2$ inner product. The objective of NMA is to find the largest amplification, $G_{c'}(t) = G(t)$, say, which the disturbances can achieve over all possible initial conditions, i.e., 
\begin{eqnarray}\label{AmplificatIon_defn}
G(t)&= & G(t, k, R):=\max_{c'_0} \Psi_{c'}(t)= \max_{c'_0} \| \Phi(t_0;t)c'_0\|_2 \nonumber \\
& = & \| \Phi(t_0; t) \| = \displaystyle \sup_{j} s_j(t),
\end{eqnarray} 
where $s_j$'s are the singular values of $\Phi(t_0; t)$, that is the eigenvalues of $\Phi(t_0; t)^* \Phi(t_0; t)$ and can be found by SVD of $\Phi(t_0; t)$. Thus, the quantities that determine the optimal amplification and the finite time behavior of energy growth are eigenvalues of $\Phi(t_0; t)^* \Phi(t_0; t)$. It must be noted that we do not calculate the $E_{c'}(t)$ explicitly, instead, we are using the classical definition of amplification magnitude  of disturbances to find the optimum amplification and the optimum perturbations by SVD. Further, there exists other possible ways to describe such transient growth instability\cite{Daniel2013, Slim2010}. For example, in the study of densdity-stratifed flow Daniel \textit{et al.} \cite{Daniel2013} modified the definition of $E_{f}(t)$ (given in Eq. \eqref{enegry}) to accommodate both concentration and velocity disturbances i.e., $E_e(t):= \int_{-\infty}^{\infty}\left[ c'^2(x,t) + u'^2(x,t) \right] \text{d} x$ (see Eq. $(3.1)$ of Daniel \textit{et al.} \cite{Daniel2013}). It was concluded that for such flow problem, the amplification measure $E_{c'}(t)$ was sufficient for their stability analysis.  In an another study of Rayleigh-B\'ernard-Maragoni convection, Doumenc \textit{et al.} \cite{Doumenc2010} observed that the velocity perturbations are slaved to the temperature perturbation when the time derivative of velocity was dropped out from their model equations. Hence they discussed the stability criteria based on the optimal amplification associated to the temperature disturbances only. Similar observation is also noted by Ward \textit{et al.} \cite{Ward2014} that the stream function is slaved to the concentration field, in the study of convection of reactive solute in porous media. Thus, following the above observations and the face that the linearized equations (Eqs. \eqref{Linearized_31} and \eqref{Linearized_32}) do not contain the explicit time derivative of velocity, the concentration disturbances are pertinent in the present problem. Hence, the optimal amplification associated to the concentration disturbances is used for the non-modal stability analysis. Below we define the growth rate and the critical time for the quantitative investigation of the transient growth of perturbations.

\begin{defn}[Stability criterion and Growth rate]
The instantaneous growth rate  $\sigma$ at time $t$ of a mode with wave number $k$ can be defined as:
\begin{equation}\label{nonmodal_growth}
\sigma=\sigma (R, k, t):= \frac{1}{G(t)}\frac{\mbox{d} G(t)}{\mbox{d}t}.
\end{equation}
\end{defn}

A given perturbation is stable if $\sigma(t)<0$ and unstable for $\sigma(t)>0$. With this definition of growth rate the critical time or the onset of instability for the given perturbation can be found.
\begin{defn} [Critical time]\label{critical-time-defn}
For a given perturbation the critical time is denoted by $t_c(k,R)$ and can be defined as the first instance of time when $\frac{\mbox{d} G(t)}{\mbox{d}t}\Big|_{t=t_c} = 0$.
\end{defn} 
As noted by Doumenc \textit{et al.}\cite{Doumenc2010} that the optimum amplification (see Eq. \eqref{AmplificatIon_defn}) has to be determined with subject to the following constraints: (a) the disturbance energy at  initial time $t_0$ is equal to unity, (b) the disturbance must satisfy the linear governing equations along with the boundary conditions in the time interval $[t_0, t]$. With this observation i.e., $G(t_0) = \| \Phi(t_0;t_0) \| = 1$, a relationship between the optimum amplification, $G(t)$, and the growth rate, $\sigma(t)$, can be obtained from Eq. \eqref{nonmodal_growth} as \cite{Tilton2013} 
\begin{equation}\label{Ampli_Growth}
G(t) = \exp \left[ \int_{t_0}^t \sigma(s) \text{d}s \right].
\end{equation}

\begin{figure*}
\centering
(a) \hspace{7cm} (b)\\
\includegraphics[width=3.in, keepaspectratio=true, angle=0]{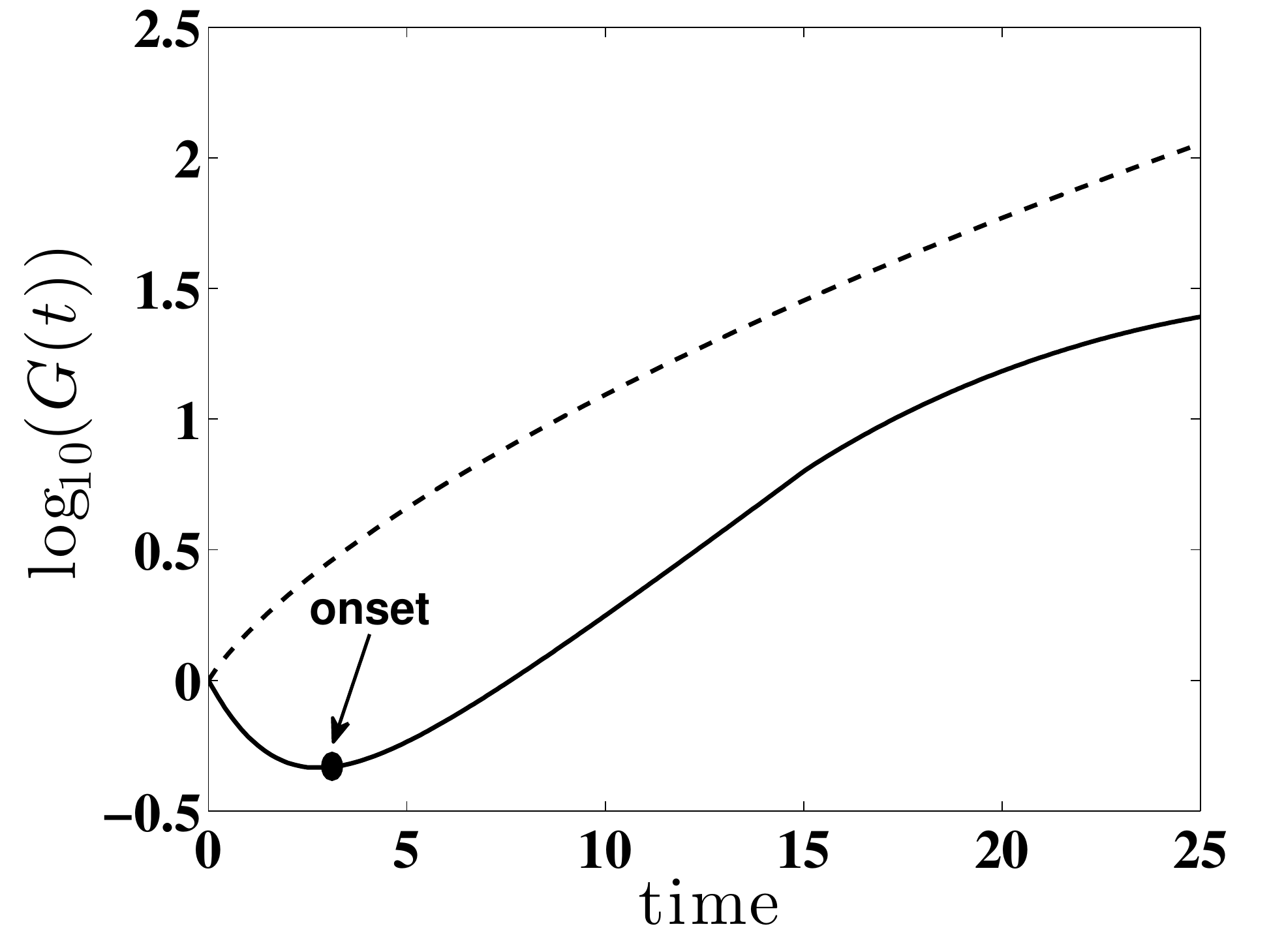}
\includegraphics[width=3.in, keepaspectratio=true, angle=0]{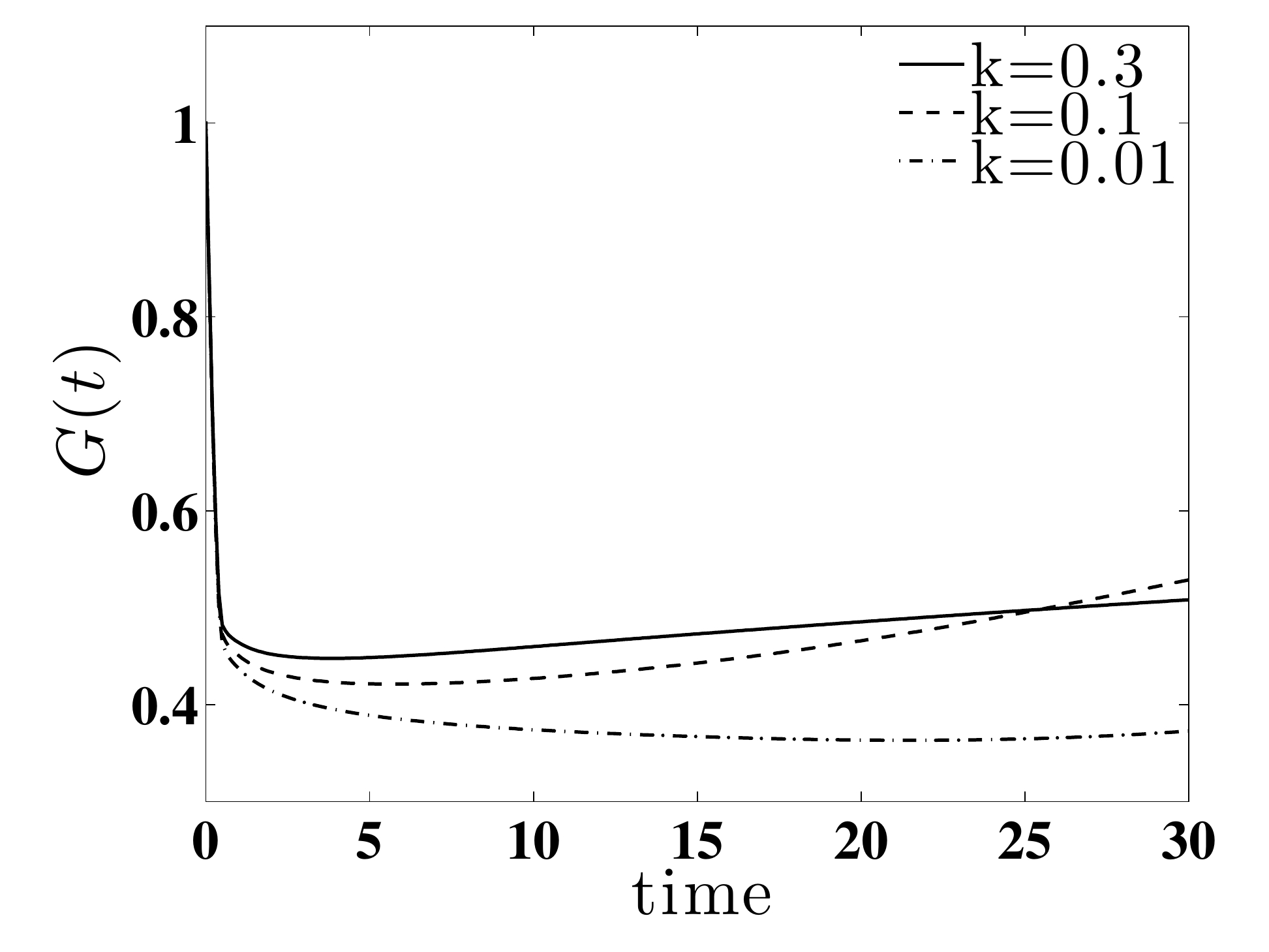}
\caption{(a)~Optimum amplification in $(x,t)$ (dashed line) and $(\xi,t)$ (continuous line) coordinate systems with $k=0.2,$ and $R=3$. In $(\xi, t)$ coordinate system, the optimal amplification $\log_{10}(G(t))$ is a non-monotonic function of time, which signifies the dominance of diffusion at early times. Also shown the onset of instability (black dot) determined by non-modal analysis in $(\xi,t)$ domain. (b)~ Optimal amplification in $(\xi, t)$ coordinate system, for $R=3$ and $k = 0.01, ~ 0.1, ~ 0.3$.}
\label{Amplification_R_3}
\end{figure*}

\section{Results and Discussion}\label{sec:NSD}


First we explain the findings of NMA, which includes the optimal  amplification of the disturbances and the spatial structure of it. We illustrate that the optimal perturbations are localized about the base state, which is contrary to the random initial perturbations that perturbs the system across the entire domain. Then, we analyse the onset of instability obtained from DNS, NMA, and other existing LSA.

\subsection{Amplification}\label{Amplificaion}
In this section, the optimal amplification $G(t)$ of a given perturbation measured from $(x,t)$ and $(\xi,t)$ coordinate systems by NMA is presented. For the base state $c_0(\xi,t)$ (see Eq. \eqref{concen_xi_t}), given a value of log-mobility ratio $R$, time $t$ and wave number $k$, the non-modal analysis determines the optimum amplification over all possible perturbations. We choose a set of wave numbers  $k = 0.01, 0.1, 0.2, 0.3$ and plotted the optimal amplification $G(t)$ as a function of time in Fig. \ref{Amplification_R_3}.

For $k=0.2$ and $R=3$, Fig. \ref{Amplification_R_3}(a) illustrates the optimal amplification $\log_{10}(G(t))$ obtained both in $(x,t)$ and $(\xi,t)$ coordinate systems. It is observed that the optimal amplification is a non-monotonic function of time in $(\xi, t)$ coordinate system.  The reason for this non-monotonicity is that at early times, the effect of diffusion is prominent which causes all the perturbations to decay. But at later time when this diffusion is weaken and convection is dominated, we observe an increase in amplification. The time when $G(t)$ starts increasing is considered as the onset of instability, which is embedded in our definition of instability criterion definition \ref{critical-time-defn}. Further, from Fig. \ref{Amplification_R_3}(a) it is observed that after the onset time $G(t)$ reaches a transient peak, which at later time must follow an exponential decay. This physical characterization of the system is appropriately captured by NMA in $(\xi,t)$ coordinate system. This is a possible reason for linear stability analysis in the self-similar coordinate system $(\xi,t)$ to give a reasonable explanation to the actual physical situation. Since the definition of amplification $G(t)$ (see Eq.\eqref{AmplificatIon_defn}) is obtained by taking all admissible perturbations, it include the mode that provides maximum growth in SS-QSSA. Thus, the curve $G(t)$ vs $t$ can be thought of as an envelope over optimal initial conditions. In Fig. \ref{Amplification_R_3}(b), it is shown that for $R=3$, $k =0.1$ and $k=0.3$, $G(t)$ is a non-monotonic function of time, except for $k=0.01$, for which $G(t)$ decays monotonically.
Thus, there exists a range of wave numbers for which transient growth is prominent. This transient growth of perturbations is consistent with our observations in Sec. \ref{subsec:Pseudo} obtained in the study of numerical and spectral abscissa. In that section we have shown that at early time there is a substantial difference between $\alpha(\underline{A})$ and $\eta(\underline{A})$.

\begin{figure}[!ht]
\centering
\includegraphics[width=3.2in, keepaspectratio=true, angle=0]{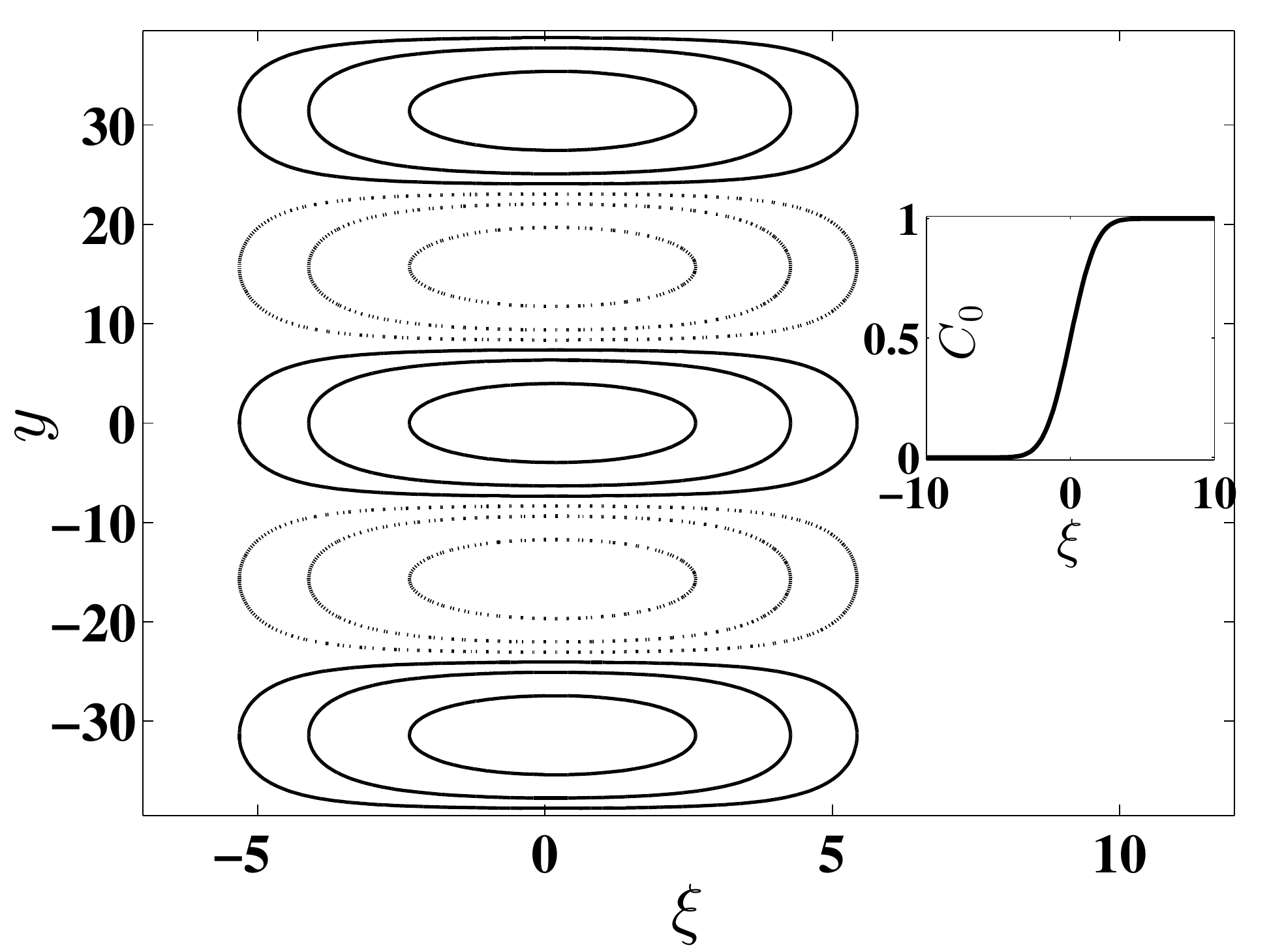}
\setlength{\abovecaptionskip}{0.1cm}
\caption{Contour plot of optimum initial perturbation at time $t=0.1$, $k=0.2$, and $R=3$ (dotted line correspond to negative contour lines and continuous line corresponds to positive contour lines). Inset shows the base state concentration $c_0$. }\label{Amplification_contour_R_3}
\end{figure}

\begin{figure*}
\centering
(a) \hspace{7cm} (b)\\
\includegraphics[width=3.in, keepaspectratio=true, angle=0]{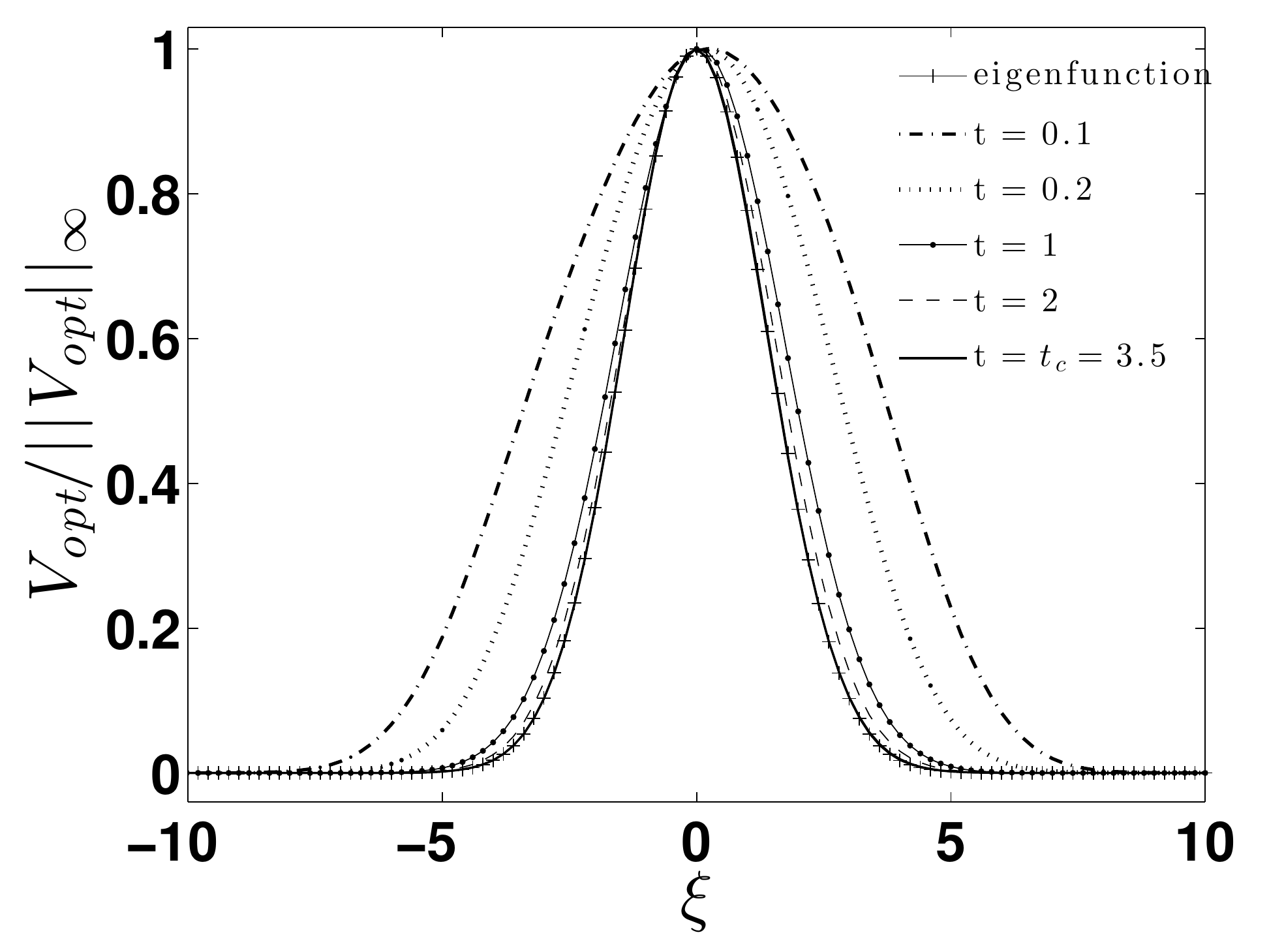}
\includegraphics[width=3.in, keepaspectratio=true, angle=0]{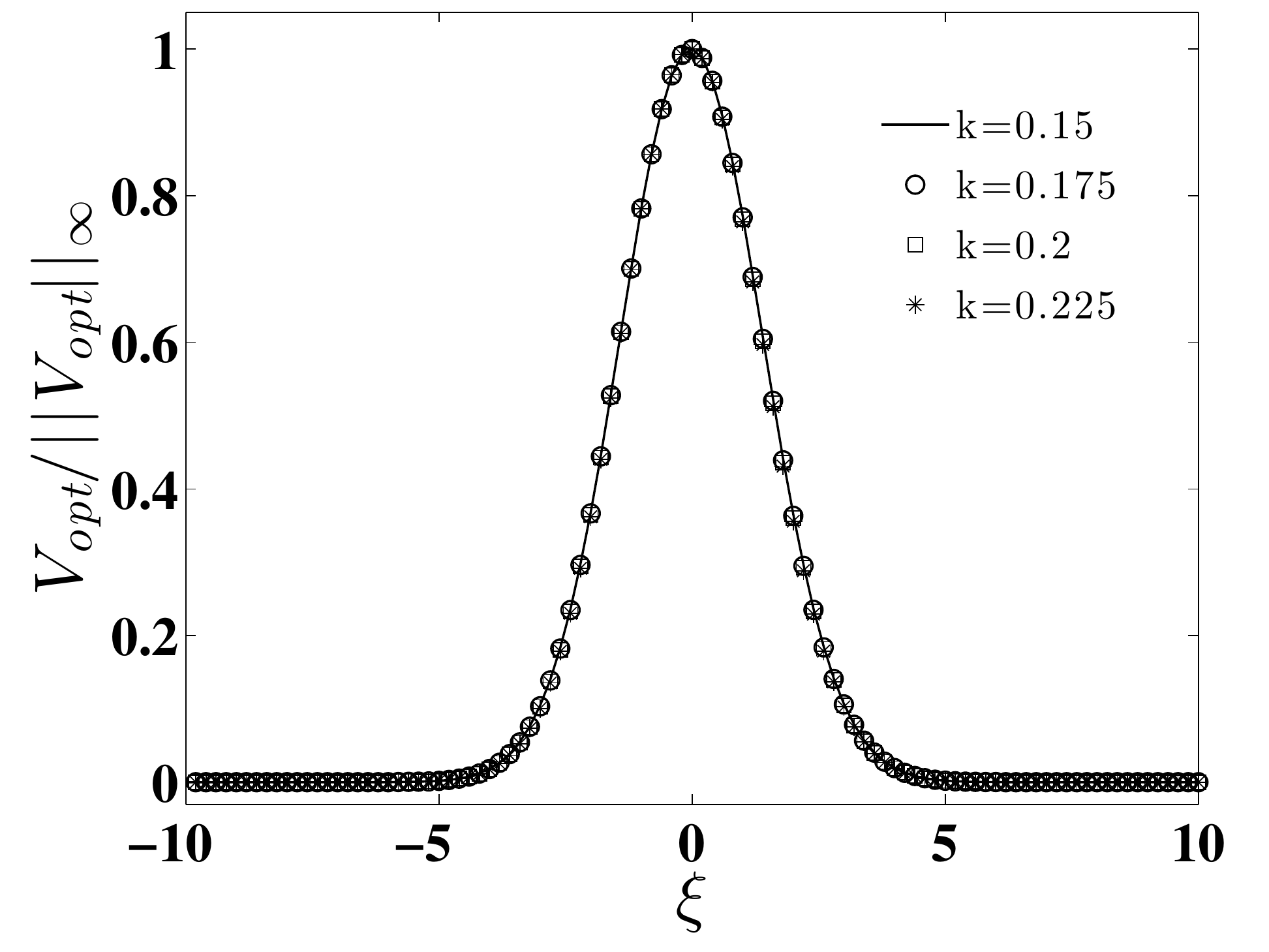}
\setlength{\abovecaptionskip}{-0.1cm}
\caption{Optimal normalized perturbation $V_{opt}$ $(a)$ for $k=0.2$ and $R=3$ at different times, with the eigenfunction calculated from SS-QSSA at $t=3.5$, $(b)$ for $R=3$ and $k=0.15, 0.175, 0.2, 0.225$ at their respective critical times calculated from definition \ref{critical-time-defn}.}\label{Optimal_evolved_IC_R_3}
\end{figure*}

We observe that the range of wave number between $0.1$ and $0.3$ dominate most, and the most unstable wave number predicted from the modal analysis \cite{Tan1986} lies within this interval. Thus, it can be claimed that the non-modal theory not only determines the transient growth, but also is a generalization of modal and IVP analyses. It is also observed when $k \to 0$ (zero), Eqs. \eqref{Linearized_31}  and \eqref{Linearized_32} with boundary conditions Eqs. \eqref{bc_1} \ and \eqref{bc_2} simplify to
\begin{eqnarray}
\left[ t\frac{\partial}{\partial t}  - \frac{\xi}{2} \frac{\partial}{\partial \xi}- \frac{\partial^2}{\partial \xi^2}\right] c'(\xi,t)&=& 0.
\end{eqnarray} 
\noindent From Fig. \ref{Amplification_R_3}(b) it is shown that unlike density fingering\cite{Daniel2013}, the amplification is a non-constant function of time for any finite wave number. This is in good agreement with the previously studied linear stability analysis of VF in the self-similar coordinate systems \cite{Ben2002, Kim2012, Satyajit2013}. 

\subsection{Optimal Perturbations}\label{Optimal_pertb}
One of the most important aspects of the non-modal analysis is to investigate the disturbances that produce maximum response during the fingering process. It is not clear from the existing linear stability analyses of miscible VF what will be the dominant perturbations at early time on the order of $t \sim t_c$. This section provides physical explanation of the transient period by examining the optimal perturbations and eigenmodes obtained from SS-QSSA. The maximal amplification $G(t)= \|\Phi(t_0, t)\|$ can be found by computing its singular value  decomposition or Schmidt decomposition \cite{Golub2007},
\begin{equation}\label{SVD_1}
\Phi(t_0, t) = \mathbf{U}_{[t_0,t]}{\sum}_{[t_0,t]}\mathbf{V^*}_{[t_0,t]},
\end{equation}  
where $\mathbf{V^*}$ represents the adjoint of the matrix $\mathbf{V}$. The largest singular value is the optimal amplification for the given perturbation i.e., the first entry of the diagonal matrix ${\sum}_{[t_0,t]}$, and the corresponding column of $\mathbf{V}$ (first right singular vector) is the optimal perturbation (initial condition), denoted by $V_{opt}$. This optimal initial condition evolves into the first column of $\mathbf{U}$ (the first left singular vector), denoted by $U_{opt}$. This is also evident from the  singular value decomposition of $\Phi(t_0, t)$, i.e., 
$\Phi(t_0, t) \vec{v}_1= s_{1,1} \vec{u}_1, \text{where} ~s_{1,1}$ is the first entry of the diagonal matrix ${\sum}_{[t_0,t]}$. Thus, the leading vectors of $\mathbf{V}$ are the useful inputs for selecting the correct representative for optimal initial conditions. Fig. \ref{Amplification_contour_R_3} shows the contour of the initial perturbation of concentration that is most amplified at $t=0.125$. As expected, the perturbation is localized within the diffusive layer. In Fig. \ref{Optimal_evolved_IC_R_3}(a) we plot the normalized optimal perturbation $V_{opt}$ for fixed $k=0.2$ at time $t=0.125, 0.25, 1, 2,$ and $4.125$. It is shown that these right singular vectors are situated around the base state and gradually resolves to the dominant eigenmode calculated from modal analysis discussed by Ben \textit{et al.} \cite{Ben2002}. In Fig.~\ref{Optimal_evolved_IC_R_3}(b) the normalized $V_{opt}$, for various wave numbers are shown at the respective critical time (see definition \ref{critical-time-defn}). Two important observations are noted from the structure of the optimal perturbations determined from non-modal theory: (i) the optimal perturbations at their respective critical times are identical to the eigenmode predicted by SS-QSSA, irrespective of any wave number chosen, and, (ii) at early times the SS-QSSA eigenmode and the optimal perturbations differ substantially, which is not shown for brevity. This reflects that although SS-QSSA captures the early times behavior better than QSSA and predicts the onset of instability, it fails to capture the transient growth of perturbations.  Hence, a physically relevant transient period for VF can be defined as $[t_0, t_c]$, where $t_0$ is the time when the time integration starts and $t_c$ is the critical time obtained from NMA. 

\section{Nonlinear Simulations}\label{sec:DNS}
In this section, we discuss the direct numerical simulations for solving the coupled nonlinear  equations that are compared with the transient growth obtained from NMA. We solve the nonlinear problem using a highly accurate pseudo-spectral method based on stream function formulation of Eqs. \eqref{cont_eqn}-\eqref{convec_diffuse}, proposed by Tan and Homsy \cite{Tan1988}. This method has been employed successfully to obtain highly accurate results for various viscous fingering problems \cite{Mishra2008}. Writing the unknown variables as, $\psi(x,y,t) - \psi_0(x,t) = \psi'(x,y,t), ~ c(x,y,t) - c_0(x,t) = c'(x,y,t)$, the stream function form of Eqs. \eqref{cont_eqn}-\eqref{convec_diffuse} in terms of the perturbation quantities can be represented by, 
\begin{eqnarray}
\label{eq:VS1}
& & \nabla^2\psi' = -R\left(\nabla\psi'\cdot\nabla(c_0 + c') + \frac{\partial c'}{\partial y}\right), \\
\label{eq:VS3}
& & \frac{\partial c'}{\partial t} + \frac{\partial \psi'}{\partial y}\left(\frac{\partial c_0}{\partial x} + \frac{\partial c'}{\partial x}\right) - \frac{\partial \psi'}{\partial x}\frac{\partial c'}{\partial y} = \nabla^2 c'.
\end{eqnarray}
 
Here the base-state flow is the same as discussed in Sec. \ref{subsec:BS}, i.e. $\psi_0 = 0$, $c_0$ is given by Eq. \eqref{base_state} and the primes correspond to their perturbation quantities (not necessarily infinitesimal). The non-dimensional width of the computational domain becomes $\mbox{Pe} = UH/D$, the P\'eclet number and the corresponding length of the domain is $A\cdot \mbox{Pe}$. Here $A = L/H$ is the aspect ratio and $L, H$ being the dimensional length and width of the computational domain, respectively. Periodic boundary conditions are applied in the transverse direction. In the longitudinal direction, we work with the periodic extension of a displacement front. In this paper, the computational domain is chosen in such a way that Pe $= 512$ and $A = 8$ with $1024 \times 128$ grid points discretizing the domain. The time integration is performed by taking time stepping $10^{-3}$.  

The simulations are performed with $t_0 = 0.01$ and initial concentration perturbation of the form,
\begin{equation}
\label{eq:IC}
c'(x,y,0) = \epsilon~\text{rand}(\cdot)\sin(ky).
\end{equation}
Here, $k$ is the wave number, $\text{rand}(\cdot)$ represents a random number between $0$ and $1$ at the diffusive interface, $\epsilon$ is the amplitude of the perturbation, which is taken to be $10^{-2}$. Assuming that the perturbations grow exponentially, the growth rate of perturbations, $\sigma_{DNS}$, can be computed at every time from,
\begin{equation}
\label{eq:GR_DNS}
\sigma_{DNS} \equiv \frac{1}{2}\frac{\text{d}[\ln(E_{DNS}(t))]}{\text{d}t}.
\end{equation}
Here, $E_{DNS}(t) \equiv \int_0^{A\cdot {\rm Pe}}\int_0^{{\rm Pe}}[c'(x,y,t)]^2\text{d}x\text{d}y$ represents an amplification measure of the perturbed concentration. The DNS results are averaged over ten different random realizations of the initial concentration perturbations, Eq.\eqref{eq:IC}. For our analysis DNS are performed for seven different wave numbers and the results obtained are discussed in Sec. \ref{QSSA_IVP_NONMODAL}. 

\subsection{Comparison of non-modal theory with SS-QSSA, IVP and DNS}\label{QSSA_IVP_NONMODAL}
In the previous sections (sec. \ref{Amplificaion} and \ref{Optimal_pertb}), it is shown that the structure of the optimal perturbation describes about the regime of transient growth. Also from the discussion of numerical and spectral-abscissa (see Sec. \ref{subsec:Pseudo}), we noted that at early times there exists significant difference between $\alpha(\underline{A}(t))$ and $\eta(\underline{A}(t))$, which is of order $\mathcal{O}(10^7)$ (see Fig. \ref{abscissa}). To explore the relevance of this optimal perturbation to the physical system, we compare the results of non-modal analysis with those obtained from SS-QSSA, IVP and DNS. 
 
\begin{figure}[!ht]
\centering
\includegraphics[width=3.2in, keepaspectratio=true, angle=0]{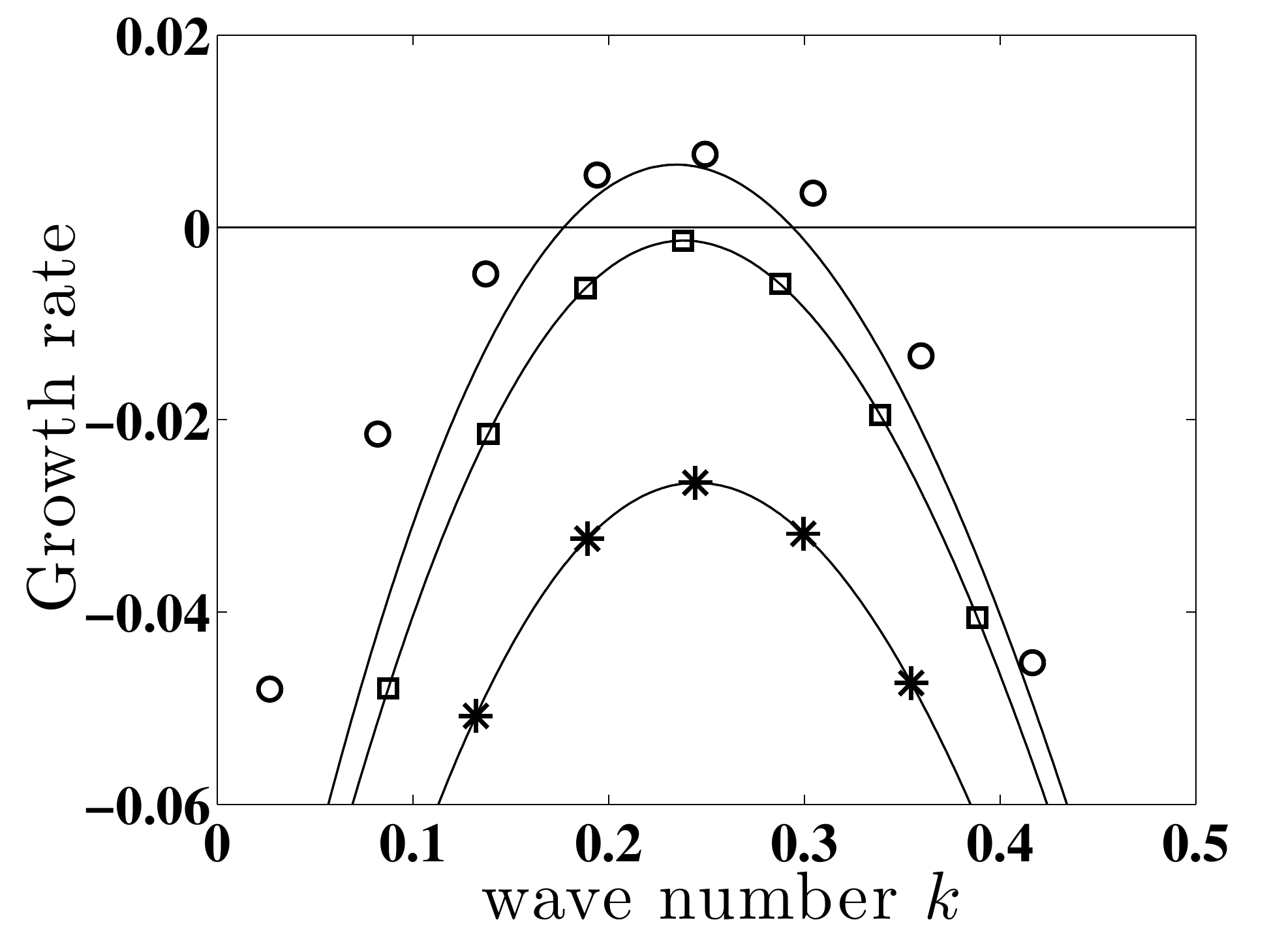}
\setlength{\abovecaptionskip}{0.1cm}
\caption{Comparison of dispersion curves obtained from SS-QSSA (line with asterisk), IVP (line with square), NMA (continuous line), and DNS (circles) for $R=3$ at time $t=3.5$.}\label{Comparisondispersion_with_QSSA}
\end{figure}

\subsubsection{Dispersion Curves and Growth rates}\label{Dispersion_Growrthrate}
\begin{figure*}
\centering
(a) \hspace{8cm} (b)\\
\includegraphics[width=3.2in, keepaspectratio=true, angle=0]{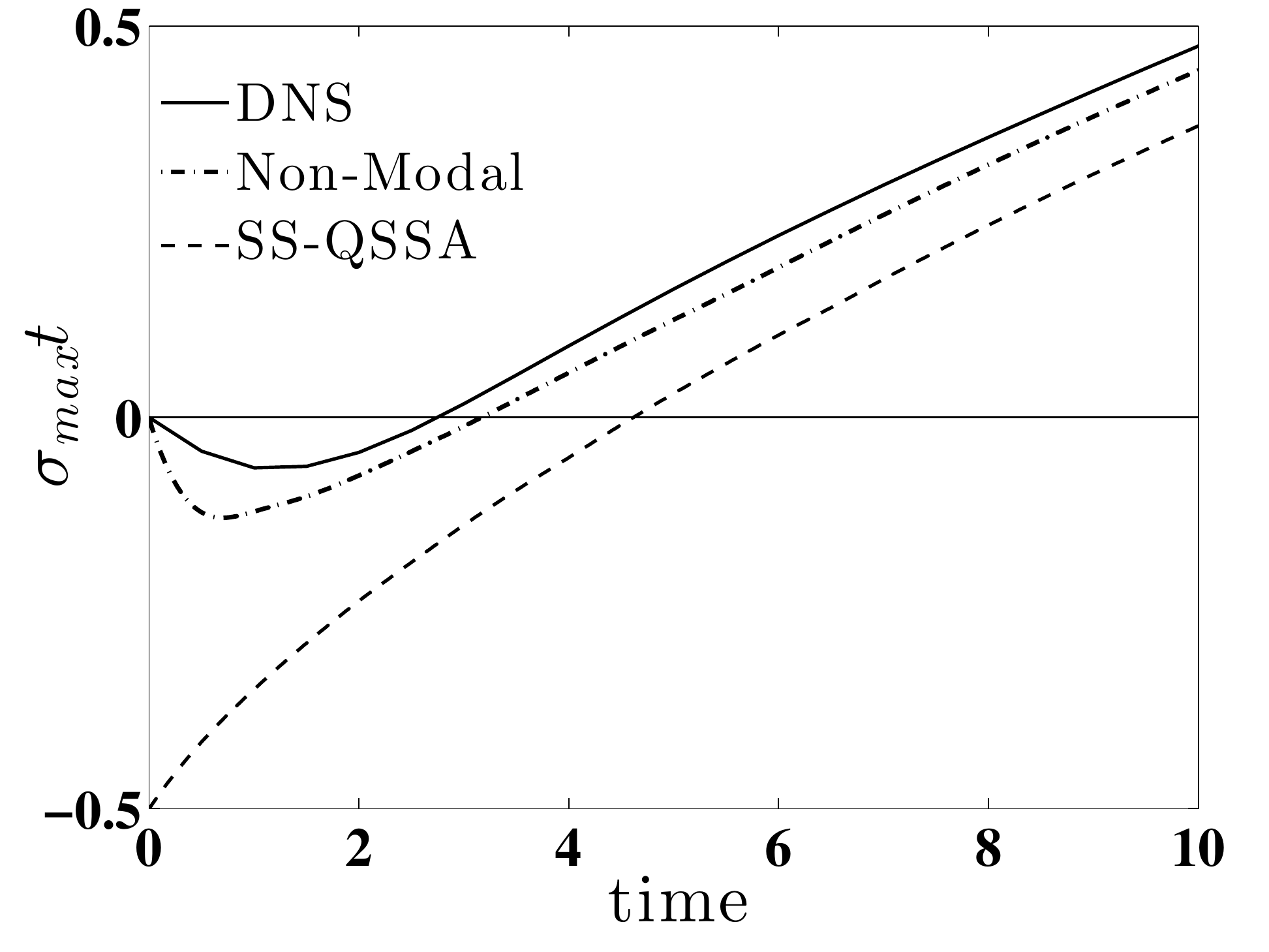}
\includegraphics[width=3.2in, keepaspectratio=true, angle=0]{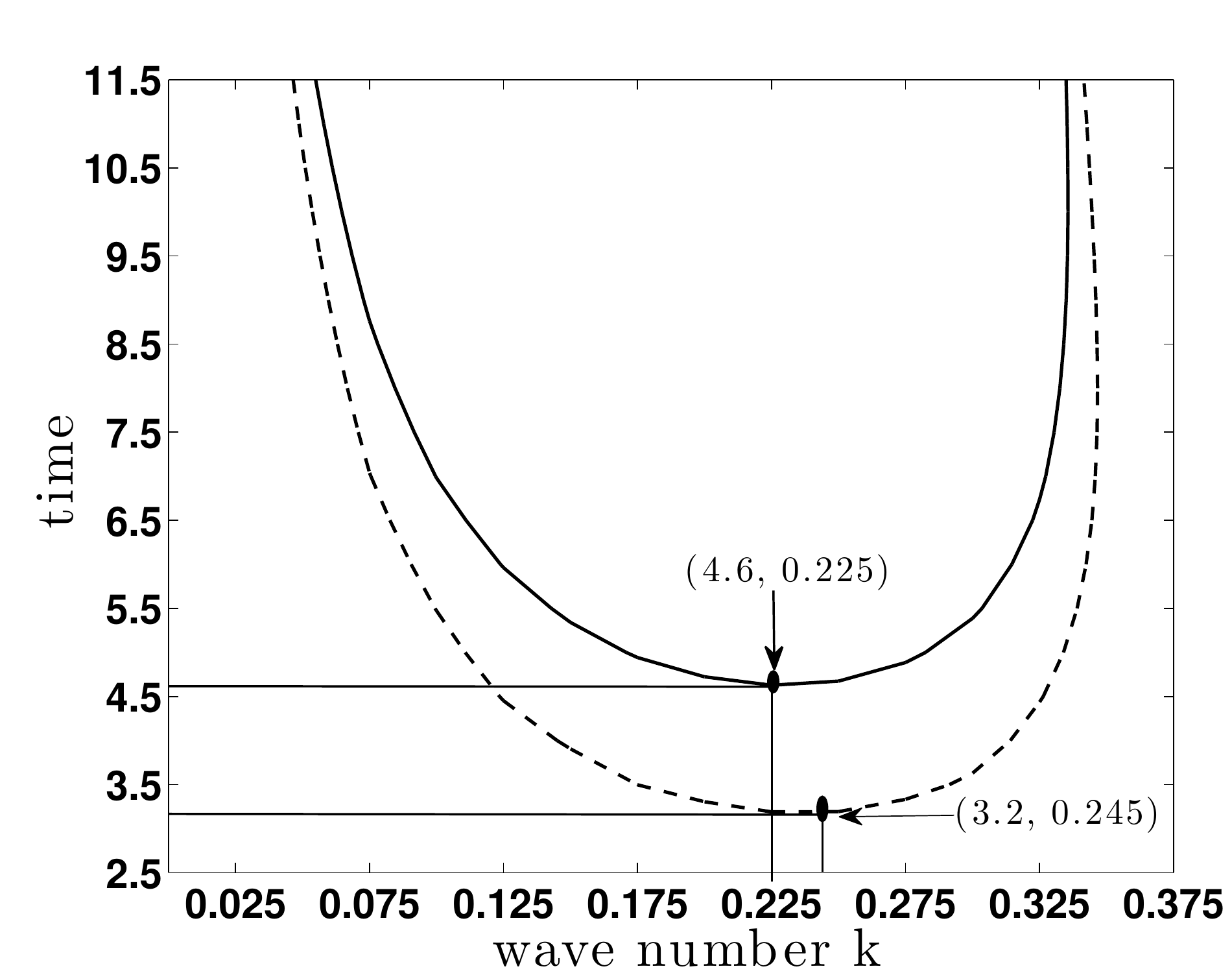}
\setlength{\abovecaptionskip}{0.1cm}
\caption{$(a)$ Optimum growth $\sigma_{max} t $ over all possible wave numbers $k$, for $R=3$, $(b)$ Neutral stability curve comparison for $R=3$: Continuous lines for SS-QSSA and dashed lines for NMA.}\label{neutral_curve_N_Optimal_growth_function}
\end{figure*}

Fig. \ref{Comparisondispersion_with_QSSA} shows the dispersion curves obtained from non-modal analysis, SS-QSSA, IVP in $(\xi,t)$ domain, and DNS at $t=3.5$. The relevance  of plotting the dispersion curves at $t=3.5$ is that it is close to the onset of fingering obtained from non-modal analysis (see Fig. \ref{neutral_curve_N_Optimal_growth_function}). It is to be noted that in Fig. \ref{Comparisondispersion_with_QSSA} (also in Fig. \ref{neutral_curve_N_Optimal_growth_function}) the time for SS-QSSA calculation  must be interpreted as diffusion time, where the base is frozen. It is evident from Fig. \ref{Comparisondispersion_with_QSSA} that non-modal analysis captures the onset of the instability  more accurately than any other linear stability methods. Although both IVP and SS-QSSA show the flow is stable, the non-modal analysis is meticulously agreeing to DNS calculation. Moreover, it is shown that the threshold wave number (the least wave number at which  $\sigma =0$) and the cut-off wave number $k_c$ (wave number for which $\sigma$ changes from positive to negative) obtained from non-modal analysis are closer to DNS than those obtained from SS-QSSA  and IVP. Thus, there exists notable transient effect of linearized non-normal matrices in VF and the present nonlinear simulations agree with the non-modal linear stability results at early times. 

\begin{figure}[!ht]
\centering
\includegraphics[width=3.2in, keepaspectratio=true, angle=0]{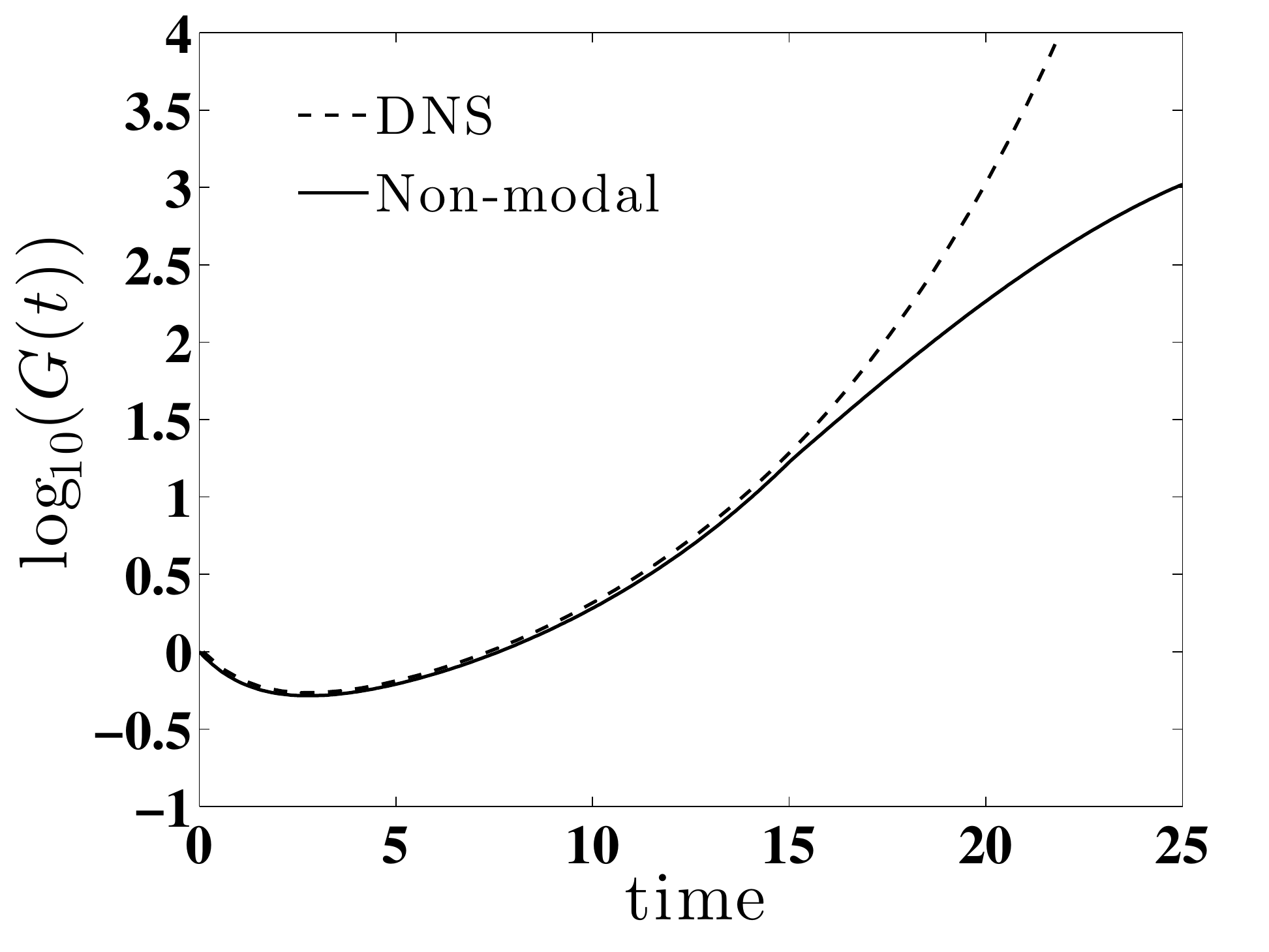}
\setlength{\abovecaptionskip}{0.1cm}
 \caption{Comparison of the amplifications from non-modal theory (continuous line) and DNS (dashed line)  for $R=3$ and $k = 0.25$.}\label{Amp_DNS_NMA}
\end{figure}

With these agreement of growth rate in NMA and DNS, we plot the amplification obtained from DNS and non-modal theory in Fig.\ref{Amp_DNS_NMA} for the wave number $k=0.25$, which amplifies most at early time. The amplification for DNS calculations are obtained using Eq. \eqref{Ampli_Growth}. It is shown that non-modal theory has an excellent agreement with DNS calculations up to $t \approx 15$. This confirms the physical relevance of the transient growth calculation from non-modal theory and the instance, when the two curves in Fig. \ref{Amp_DNS_NMA} deviate from each other can be marked as the threshold of nonlinear convection.

\subsubsection{Optimal growth and dominant wave numbers}\label{Max_waveno_Optimal_Growrth}
In an experiment or in any physical process a perturbation consists of combination of  different wave numbers. Therefore, it is important to  calculate the onset of instability by considering all possible wave numbers. This will characterize what will be the optimum growth at a given time. Fig. \ref{neutral_curve_N_Optimal_growth_function}(a) depicts that the onset time obtained from modal, non-modal, and DNS are $4.6,\; 3.2, \text{and}\:  2.8$, respectively. One of the important observations from optimal growth calculation is that the effect of the diffusion at early time is well described by non-modal and DNS, whereas SS-QSSA fails to observe this physical phenomenon effectively. The reason that NMA explains the physical phenomenon appropriately is  attributed to the fact that at early times the linearized matrix $\underline{A}(t)$ is highly non-normal, which is not reported in earlier linear stability analyses. Since at very early stage of the evolution of the perturbations, the growth rate in NMA and SS-QSSA are all negative, it does not affect the instability. 

Thus, at early stage ($t \ll 1$) in the linear regime optimal growth is damped before it starts to grow. The instance when both the convection and the diffusion terms are balanced, i.e. when $\sigma = 0$, is termed as the neutral stability condition. It is often required and useful to identify the unstable and stable regimes precisely. Fig. \ref{neutral_curve_N_Optimal_growth_function}(b) represents the neutral stability curve determined by modal and non-modal theories. Again, the qualitative features remain the same, but there is a quantitative difference at early times and hence the stable regime identified by modal analysis becomes unstable for NMA. The lowest point on the curves corresponds to the critical time and critical wave number. This clearly depicts that although the critical wave numbers obtained from both modal and non-modal analyses are the same, the critical time is smaller in non-modal theory than modal analysis.

\section{Conclusions}
A novel non-modal linear stability analysis is presented to study the transient growth of perturbation in miscible VF. Recent works\cite{Matar1999a, Rapaka2008, Daniel2013} suggested that in the unsteady base flow problem, the transient response of the perturbations can be significant. The existing literature does not discuss the transient effect of time dependent base state in the case of viscosity unstable problems. Our approach of non-modal analysis of VF is based on finding the propagator matrix and singular value decomposition method \cite{Rapaka2008, Rapaka2009}. This method predicts the dominant perturbations that experience the maximum amplification within the linear regime. The present linear stability analysis based on sounder mathematical basis have two most important features: (i) it determines the optimal amplification of the perturbation that is imperceptible in both SS-QSSA and IVP analyses, and (ii) it identifies the physical mechanism of the  VF instability, which is in very good agreement with DNS results. Additionally, the structure of the initial perturbations which lead to the optimal amplification is also illustrated, along with all the qualitative information of flow instability that SS-QSSA and IVP together provides. Unlike the random perturbations which disturb the system everywhere, the initial perturbations obtained from NMA are localized within the diffusive zone, which is in agreement with the spectral analysis of Ben \textit{et al.} \cite{Ben2002}. Although matrix differential method is used in the present study, the adjoint-looping method \cite{Corbett2000, Doumenc2010, Daniel2013} can be an useful alternate way to study the transient growth. Further, this classical non-modal analysis can be helpful to find the suitable time and length scales for the instability, the structure of most amplified initial condition in many important hydrodynamic instability problems, which deals with the unsteady base state. For example, the effects of viscosity contrast on buoyantly unstable miscible fluids in a porous medium, in understanding the effect of precipitation reactions during flow displacements in porous media in the context of CO$_2$ sequestration techniques\cite{Nagatsu2014, Satyajit2015}. 

\section*{Acknowledgments}\label{sec:acknowledgment}
Authors are thankful to the Department of Science and Technology, Government of India for its financial support during preparation of this manuscript. S.P. gratefully acknowledges the National Board for Higher Mathematics, Government of India for the financial support through a Ph.D. fellowship. 


\end{document}